\documentclass[article,nojss]{jss}
\usepackage{amsmath}
\usepackage{amssymb}
\usepackage{graphicx}
\usepackage{mathtools}
\usepackage{tabularx}
\usepackage{algorithm}
\usepackage{algpseudocode}
\usepackage{rotating}
\usepackage{booktabs}
\usepackage{multirow}
\DeclareMathOperator*{\argmin}{argmin}
\DeclareMathOperator*{\argmax}{arg\,max}

\author{Samuel Watson\\University of Birmingham}
\title{Generalised Linear Mixed Model Analysis, Maximum Likelihood Fitting, Prediction, Simulation, and Optimal Design in \proglang{R}}

\Plainauthor{Samuel Watson} 
\Plaintitle{Generalised Linear Mixed Model Analysis, Maximum Likelihood Fitting, Prediction, Simulation, and Optimal Design in R} 
\Shorttitle{glmmr packages for R} 

\Abstract{
   We describe the \proglang{R} package \pkg{glmmrBase} and an extension \pkg{glmmrOptim}. \pkg{glmmrBase} provides a flexible approach to specifying, fitting, and analysing generalised linear mixed models. We use an object-orientated class system within \proglang{R} to provide methods for a wide range of covariance and mean functions relevant to multiple applications including cluster randomised trials, cohort studies, spatial and spatio-temporal modelling, and split-plot designs. The class generates relevant matrices and statistics and a wide range of methods including full maximum likelihood estimation of generalised linear mixed models using stochastic Maximum Likelihood, Laplace approximation, power calculation, standard error corrections, random effect and broader data simulation, and access to relevant calculations. The package includes Markov Chain Monte Carlo simulation of random effects, sparse matrix methods, and other functionality to support efficient estimation. The \pkg{glmmrOptim} package implements a set of algorithms to identify c-optimal experimental designs where observations are correlated and can be specified using the generalised linear mixed model classes. Several examples and comparisons to existing packages are provided to illustrate use of the packages.
}
\Plainkeywords{Markov Chain Monte Carlo Maximum Likelihood, Stochastic Approximation Expectation Maximisation, Generalised Linear Mixed Model, R, C++, c-Optimal Experimental design} 

\Address{
  Samuel Watson\\
  Institute for Applied Health Research\\
  University of Birmingham \\
  Birmingham, UK\\
  E-mail: \email{S.I.Watson@bham.ac.uk}
}

\begin{document}


\section{Introduction}

Generalised linear mixed models (GLMM) are a highly flexible class of statistical models that incorporate both `fixed' and `random' effects. GLMMs permit the incorporation of latent effects and parameters and allow for complex covariance structures. For example, they are widely used in the analysis of: clustered data, such as from cluster randomised trials, to capture latent cluster means; cohort studies to incorporate temporal correlation between observations on individuals; or, in geospatial statistical models as the realisation of a Gaussian process used to model a latent spatial or spatio-temporal surface. Their use in such a wide variety of statistical applications means there exist several packages and libraries for software including \proglang{R}, Stata, and SAS to provide relevant calculations for study design, model fitting, and other analyses. 

For many types of analysis, such as a power calculation for a cluster trial, there exist multiple different packages each implementing a set of specific models. Users may therefore be required to use multiple packages with different interfaces for a single analysis. A more general and flexible system that provides a wide range of functionality for this model type and that  permits users to add and extend functionality may therefore simplify statistical workflows and facilitate more complex analyses.

In this article, we describe the \proglang{R} package \pkg{glmmrBase} and an extension \pkg{glmmrOptim} for the \proglang{R} programming language. These packages provide a general framework for GLMM specification, including non-linear functions of parameters and data, with calculation of relevant matrices, statistics, and other functions designed to provide useful analyses for a large range of model specifications and to support implementation of other GLMM related software in \proglang{R}. The aim of this \proglang{R} package (and its underlying \proglang{C++} library) was to provide several features altogether not available in other software packages: MCMC Maximum likelihood model fitting, run-time specification of non-linear fixed effect forms for GLMM models, GLMM model fitting with flexible, easy-to-specify covariance functions, data simulation and model summary features, easy access to relevant matrices and calculations, and support for extensions and algorithms including optimal design algorithms. We summarise and compare existing software that provides functionality in these areas where relevant in each section. 

\subsection{Generalised linear mixed models}
\label{subsec: glmm_framework}
A generalised linear mixed model (GLMM) has the linear predictor for observation $i$
\begin{equation*}
    \eta_i = g(\mathbf{x}_i,\boldsymbol{\beta}) + \mathbf{z}_i \mathbf{u}
\end{equation*}
where $\mathbf{x}_i$ is the $i$th row of matrix $X$, which is a $n \times P$ matrix of covariates, $\boldsymbol{\beta}$ is a vector of parameters, $\mathbf{z}_i$ is the $i$th row of matrix $Z$, which is the $n \times Q$ ``design matrix'' for the random effects, and $\mathbf{u} \sim N(0,D)$, where $D$ is the $Q \times Q$ covariance matrix of the random effects terms that depends on parameters $\boldsymbol{\theta}$. In this article we use bold lower case characters, e.g. $\mathbf{x}_i$ to represent vectors, normal script lower case, e.g. $\beta_1$, to represent scalars, and upper case letters, e.g. $X$, to represent matrices.

The model is then
\begin{equation*}
y_i \sim G(h(\eta_i);\phi)
\end{equation*}
$\mathbf{y}$ is a $n$-length vector of outcomes with elements $y_i$, $G$ is a distribution, $h(.)$ is the link function such that $\mu_i = h(\eta_i)$ where $\mu_i$ is the mean value, and $\phi$ is an additional scale parameter to complete the specification. 

When $G$ is a distribution in the exponential family, we have:
\begin{align}
\begin{split}
    f_{y|\mathbf{u}}(y_i|\mathbf{u}, \beta, \phi) &= \exp{(y_i\eta_i - c(\eta_i))/a(\phi) + d(y,\phi)} \\
    \mathbf{u} & \sim f_{\mathbf{u}}(\mathbf{u}|\theta)
    \end{split}
\end{align}
The likelihood of this model is given by:
\begin{equation}
\label{eq:lik1}
    L(\beta,\phi,\theta|\mathbf{y}) = \int \prod_{i=1}^n f_{y|\mathbf{u}}(y_i|\mathbf{u}, \beta, \phi)f_{\mathbf{u}}(\mathbf{u}|\theta) d \mathbf{u}
\end{equation}
The likelihood (\ref{eq:lik1}) generally has no closed form solution, and so different algorithms and approximations have been proposed to estimate the model parameters. We discuss model fitting in Section \ref{sec:mcml}. Where relevant we represent the set of all parameters as $\boldsymbol{\Theta} = (\boldsymbol{\beta}, \phi, \boldsymbol{\theta})$.

\section{A GLMM Model in glmmrBase}
The \pkg{glmmrBase} package defines a \code{Model} class using the \pkg{R6} class system \citep{r6}, which provides an encapsulated object orientated programming system for \pkg{R}. Most of the functionality of the package revolves around the \code{Model} class, which interfaces with underlying \proglang{C++} classes. A non-exhaustive list of the functions and objects calculated by or a member of the \code{Model} class is shown in Table \ref{tab:model}. The class also contains two subclasses, the \code{covariance} and \code{mean} classes, which handle the random effects and linear predictor, respectively. \pkg{R6} classes allow encapsulating classes to `share' class objects, so that a single \code{covariance} object could be shared by multiple \code{Model} objects. \pkg{R6} also provides familiar object-orientated functionality, including class inheritance, so that new classes can be created that inherit from the \code{Model} class to make use of the range of functions. Within \proglang{R}, we use the \pkg{Matrix} package for matrix storage and linear algebra for operations, and the Eigen \proglang{C++} library for linear algenra functionality in \proglang{C++}. Linkage between \proglang{R} and \proglang{C++} is provided through the \proglang{Rcpp} and \proglang{RcppEigen} packages \citep{Eddelbuettel2011}. We describe the base functionality and model specification in this section, and then describe higher-level functionality including model fitting in subsequent sections. The underlying \proglang{C++} library is header-only and so can be imported into other projects.

\begin{table}[]
    \centering
    \small
    \begin{tabular}{p{0.12\linewidth}p{0.25\linewidth}|p{0.4\linewidth}|p{0.1\linewidth}}
    \toprule
    \multicolumn{2}{l|}{\textbf{Method}} & \textbf{ Description} & \textbf{Sec.} \\
    \midrule
       \multirow{9}{*}{\code{covariance}}  & \code{D} & The matrix $D$ & \ref{subsec:covcalc}\\
       & \code{Z} & The matrix $Z$ & \ref{sec:matZ}\\
         & \code{chol\_D()} & Cholesky decomposition of $D$ & \ref{subsec:covcalc}\\
         & \code{log\_likelihood()} & Multivariate Gaussian log-likelihood with zero mean and covariance $D$ & \ref{subsec:covparest}\\
         & \code{simulate\_re()} & Simulates a vector $\mathbf{u}$ & \ref{sec:datasim}\\
         & \code{sparse()} & Choose whether to use sparse matrix methods & \ref{subsec:covcalc} \\
         & \code{parameters} & The parameters $\theta$\\
         & \code{formula} & Random effects formula & \ref{subsec:cov_fun} \\
         & \code{update\_parameters()} & Updates $\theta$ and related matrices \\
         & \code{hsgp()} & HSGP approximation parameters & \ref{subsec:cov_fun} \\
         & \code{nngp()} & NNGP approximation parameters & \ref{subsec:cov_fun} \\
         \midrule
         \multirow{6}{*}{\code{mean}} & \code{X} & The matrix X &  \\
         & \code{parameters} & The parameters $\beta$ \\
         & \code{offset} & The optional model offset \\
         & \code{formula} & The fixed effects formula used to create $X$ & \ref{subsec:mean_fun} \\
         & \code{linear\_predictor()} & Generates $X\beta$ plus offset \\
        & \code{update\_parameters()} & Updates $\beta$ and related matrices \\
        \midrule
        \multicolumn{2}{l|}{\code{family}} &  A \proglang{R} family object \\
        \multicolumn{2}{l|}{\code{var\_par}} & An optional scale parameter\\
        \multicolumn{2}{l|}{\code{fitted()}} & Full linear predictor \\
        \multicolumn{2}{l|}{\code{predict()}} & Predictions from the model at new data values & \ref{sec:predict} \\
        \multicolumn{2}{l|}{\code{sim\_data()}} & Simulates data from the model & \ref{sec:datasim} \\
        \multicolumn{2}{l|}{\code{Sigma()}} & Generates $\Sigma$ (or an approximation) & \ref{sec:approxsigma} \\
        \multicolumn{2}{l|}{\code{information\_matrix()}} & The information matrix & \ref{sec:approxsigma}, \ref{sec:LA} \\
        \multicolumn{2}{l|}{\code{sandwich()}} & Robust sandwich matrix & \ref{subsec:se} \\
        \multicolumn{2}{l|}{\code{small\_sample\_correction()}} & Bias-corrected variance-covariance matrix of $\hat{\beta}$ & \ref{subsec:se} \\
        \multicolumn{2}{l|}{\code{box()}} & Inferential statistics for the modified Box correction & \ref{subsec:se} \\
        \multicolumn{2}{l|}{\code{marginal()}} & Marginal effects of covariates & \ref{subsec:se} \\
        \multicolumn{2}{l|}{\code{partial\_sigma()}} & Matrices $\partial \Sigma/\partial \theta$ and $\partial^2 \Sigma/\partial \theta_i\partial \theta_j$ & \ref{subsec:se} \\
        \multicolumn{2}{l|}{\code{use\_attenutation()}} & Option for improving approximation of $\Sigma$ & \ref{sec:approxsigma} \\
        \multicolumn{2}{l|}{\code{power()}} & Estimates the power & \ref{subsec:power} \\
        \multicolumn{2}{l|}{\code{MCML()}} & Markov Chain Monte Carlo Maximum Likelihood model fitting & \ref{sec:mcml}  \\
        \multicolumn{2}{l|}{\code{LA()}} & Maximum Likelihood model fitting with Laplace approximation & \ref{sec:LA}  \\
        \multicolumn{2}{l|}{\code{mcmc\_sample()}} & Sample $\mathbf{u}$ using MCMC & \ref{sec:mcmc}  \\
        \multicolumn{2}{l|}{\code{w\_matrix()}} & Returns $diag(W)$ & \ref{sec:approxsigma}  \\
        \multicolumn{2}{l|}{\code{dh\_deta()}} & Returns $\partial h^{-1}(\eta)/\partial \eta$ & \ref{sec:approxsigma}  \\
        \multicolumn{2}{l|}{\code{calculator\_instructions()}} & Prints the calculation instructions for the linear predictor & \ref{subsec:mean_fun}  \\
        \multicolumn{2}{l|}{\code{log\_gradient()}} & Returns either $\nabla_{\mathbf{u}}L(\beta,\phi,\theta\vert \mathbf{y})$ or $\nabla_{\beta}L(\beta,\phi,\theta\vert \mathbf{y})$ & \ref{sec:mcmc}  \\
         \bottomrule
    \end{tabular}
    \caption{A non-exhaustive list of the publicly available methods and objects in the \code{Model} class. \code{Model} also contains two sub-classes \code{covariance} and \code{mean} whose respective methods are also shown. The column \textbf{Sec.} shows the relevant section of the article detailing the methods.}
    \label{tab:model}
\end{table}

An example call to generate a new \code{Model} is:
\begin{CodeChunk}
\begin{CodeInput}
model <- Model$new(formula = ~ factor(t) - 1 + (1|gr(j)*ar1(t)),
                   data = data,
                   covariance = c(0.25,0.8),
                   mean = c(rep(0,5)),
                   family = gaussian())
\end{CodeInput}
\end{CodeChunk}
We discuss each of the elements of this call in turn. 

\subsection{Data Generation Tools}
\label{subsec:other_fun}
We introduce methods to generate data for hierarchical models and blocked designs. As we show in subsequent examples, `fake' data generation is typically needed to specify a particular data structure at the design stage of a study, but can be useful in other circumstances. \citet{Nelder1965} suggested a simple notation that could express a large variety of different blocked designs. The notation was proposed in the context of split-plot experiments for agricultural research, where researchers often split areas of land into blocks, sub-blocks, and other smaller divisions, and apply different combinations of treatments. However, the notation is useful for expressing a large variety of experimental designs with correlation and clustering, including cluster trials, cohort studies, and spatial and temporal prevalence surveys. We have included the function \code{nelder()} in the package that generates a data frame of a design using Nelder's notation. 

There are two operations:
\begin{enumerate}
\item \code{>} (or $\to$ in Nelder's notation) indicates ``clustered in''.
\item \code{*} (or $\times$ in Nelder's notation) indicates a crossing that generates all combinations of two factors.
\end{enumerate}

The function takes a formula input indicating the name of the variable and a number for the number of levels, such as \code{abc(12)}. So for example \code{~cl(4) > ind(5)} means in each of five levels of \code{cl} there are five levels of \code{ind}, and the individuals are different between clusters. Brackets are used to indicate the order of evaluation. Some specific examples are illustrated in Table \ref{table:notation_example}.

\begin{table}
\centering
\begin{tabular}[t]{p{5cm}|p{8cm}}
\hline
\textbf{Formula} & \textbf{Meaning} \\
\hline
\code{~person(5) * time(10)} & A cohort study with five people, all observed in each of ten periods \code{time} \\
\hline
\code{~(cl(4) * t(3)) > ind(5)} & A repeated-measures cluster study with four clusters (labelled \code{cl}), each observed in each time period \code{t} with cross-sectional sampling and five individuals (labelled \code{ind}) in each cluster-period.\\
\hline
\code{~(cl(4) > ind(5)) * t(3)} & A repeated-measures cluster cohort study with four clusters (labelled \code{cl}) with five individuals per cluster, and each cluster-individual combination is observed in each time period \code{t}.\\
\hline
\code{~((x(100) * y(100)) > hh(4)) * t(2)}& A spatial-temporal grid of 100x100 and two time points, with 4  households per spatial grid cell.\\
\hline
\end{tabular}
\caption{\label{table:notation_example}Examples of formulae for the \code{nelder()} function}
\end{table}

Use of this function produces a data frame:
\begin{CodeChunk}
\begin{CodeInput}
data <- nelder(~(j(4) * t(5)) > i(5))
head(data)
\end{CodeInput}
\begin{CodeOutput}
>   j t i
> 1 1 1 1
> 2 1 1 2
> 3 1 1 3
> 4 1 1 4
> 5 1 1 5
> 6 1 2 6
\end{CodeOutput}
\end{CodeChunk}

The data frame shown above may represent, for example, a cluster randomised study with cross-sectional sampling. Such an approach to study design assumes the same number of each factor for each other factor, which is not likely adequate for certain study designs. We may expect unequal cluster sizes, staggered inclusion/drop-out, and so forth, and so a user-generated data set would instead be required. Certain treatment conditions may be specified with this approach including parallel trial designs, stepped-wedge implementation, or factorial approaches by specifying a treatment arm as part of the block structure.

\subsection{Linear Predictor} 
\label{subsec:mean_fun}
The package allows for flexible specification of $g(\mathbf{x_i},\boldsymbol{\beta})$. Standard model formulae in \proglang{R} can be used, for example \code{formula = ~ factor(t) - 1}, the package also allows a wide range of specifications including non-linear functions and naming parameters in the function, including allowing the same parameter to appear in multiple places. For example, the function $\beta_0 + \beta_{int}\texttt{int}_i + \beta_1\exp(\beta_2\texttt{x}_i)$ can be specified as \code{formula = ~ int + b\_1*exp(b\_2*x)}. Names of data are assumed to be multiplied by a parameter as in the standard linear predictor \code{~ x} would produce $\beta_xx$. To include data without a parameter, it can be wrapped in brackets such as \code{~(x)}, which can also provide a way of specifying offsets, although this can also be achieved by directly specifying the offset.

\begin{figure}
    \centering
    \includegraphics[width = \textwidth]{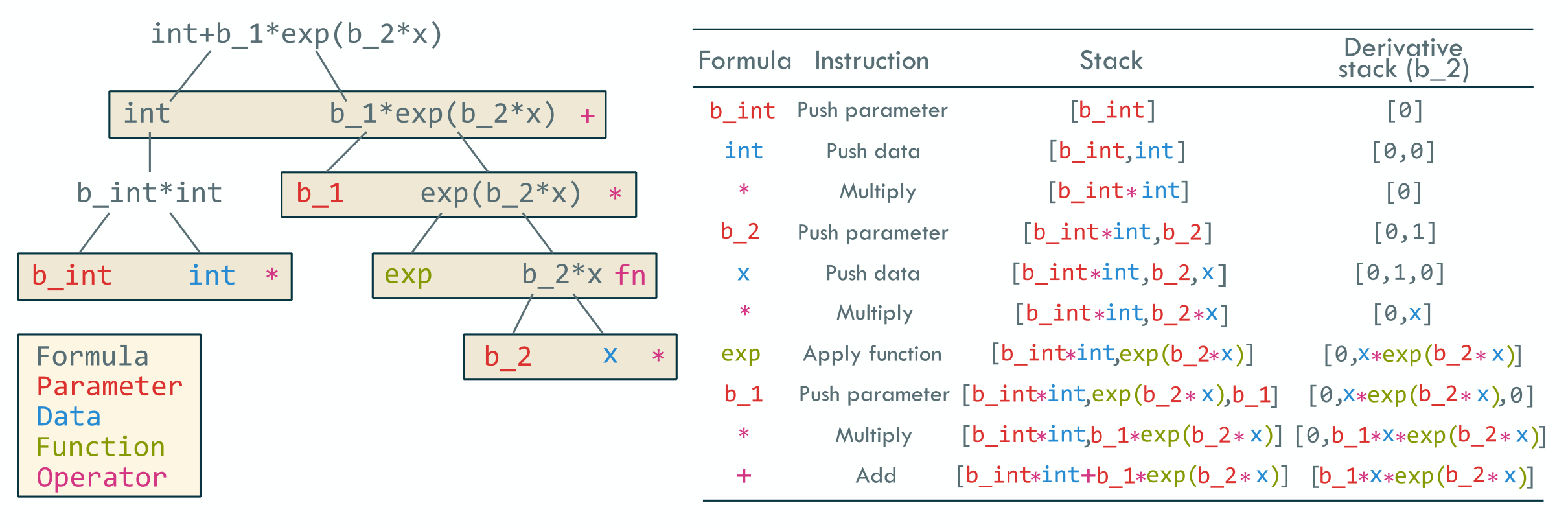}
    \caption{Illustrative example of formula parsing and calculation including autodifferention with respect to parameter \code{b\_1}}
    \label{fig:formparse}
\end{figure}

We use a recursive, operator precedence algorithm to parse the formula and generate a reverse Polish notation representing the formula. The algorithm first splits the formula at the first `+' or `-' symbol outside of a bracket, then if there are none, at a multiply (`*') or divide (`/') symbol outside of a bracket, then a power (`\^ ') symbol, and finally brackets, until a token is produced (e.g. \code{b\_2} or \code{exp}). The resulting token is first checked against available functions, then if it is not a function it is checked against the column names of the data, and if it is not in the data it is considered the name of a parameter. The resulting tree structure can then be parsed up each of its branches to produce an instruction set. Figure \ref{fig:formparse} shows an example of the formula parsing algorithm, and calculation of the value of the formula. This scheme also allows for an auto-differention approach to obtain the first and second derivatives where required. 

\label{subsec:cov_fun}
\subsection{Covariance specification}
We adapt and extend the random effects specification approach of the popular \proglang{R} package \pkg{lme4} \citep{lme4} and other related packages, building on the flexible formula parsing and calculation described above, to allow for relatively complex structures through the specification of a covariance function (see Section \ref{sec:mcml} for a summary of related packages). A covariance function is specified as an additive formula made up of components with structure \code{(z|f(j))}. The left side of the vertical bar specifies the covariates in the model that have a random effects structure; using \code{(1|f(j))} specifies a `random intercept' model. The right side of the vertical bar specifies the covariance function \code{f} for that term using variable \code{j} named in the data. Multiple covariance functions on the right side of the vertical bar are multiplied together, i.e., \code{(1|f(j)*g(t))}. The currently implemented functions (as of version 0.6.1) are listed in Table~\ref{table:cov_fun}.

\begin{table}
\centering
\small
\begin{tabular}[t]{l|l|l|l}
\hline
\textbf{Function} & $Cov(x_i,x_{i'})$ & $\theta$ & \textbf{Code}\\
\hline
Group membership & $\theta_1^2 \mathbf{1}(x_i = x_{i'})$ & $\theta_1 > 0$ & \code{gr(x)}\\
Exponential & $\theta_1 \text{exp}(- \vert x_i - x_{i'}\vert / \theta_2 )$ &$\theta_1,\theta_2 > 0$& \code{fexp(x)}\\
& $\text{exp}(- \vert x_i - x_{i'}\vert /\theta_1)$&$\theta_1 > 0$ & \code{fexp0(x)}\\
Squared Exponential & $\theta_1 \text{exp}(- (\vert x_i - x_{i'}\vert / \theta_2)^2)$ &$\theta_1,\theta_2 > 0$& \code{sqexp(x)}\\
& $\text{exp}(-( \vert x_i - x_{i'}\vert/\theta_1)^2 )$ &$\theta_1 > 0$& \code{sqexp0(x)}\\
Autoregressive order 1 & $\theta_1^{\vert x_i - x_{i'} \vert}$ &$0 < \theta_1 < 1$& \code{ar(x)}\\
Bessel & $K_{\theta_1}(x)$ & $\theta_1$ > 0 & \code{bessel(x)}\\
Matern & $\frac{2^{1-\theta_1}}{\Gamma(\theta_1)}\left( \sqrt{2\theta_1}\frac{x}{\theta_2} \right)^{\theta_1} K_{\theta_1}\left(\sqrt{2\theta_1}\frac{x}{\theta_2})\right)$ & $\theta_1,\theta_2 > 0$ & \code{matern(x)} \\
\textit{Compactly supported*} &&& \\
Truncated Power 2 & $\theta_1(1 - \vert x_i - x_{i'}\vert^{\theta_2})^2$ & $\theta_1 > 0$, $0 < \theta_2 \leq 2$& \code{truncpow2(x)} \\
Truncated Power 3 & $\theta_1(1 - \vert x_i - x_{i'}\vert^{\theta_2})^3$ & $\theta_1 > 0$, $0 < \theta_2 \leq 2$& \code{truncpow3(x)} \\
Truncated Power 4 & $\theta_1(1 - \vert x_i - x_{i'}\vert^{\theta_2})^4$ & $\theta_1 > 0$, $0 < \theta_2 \leq 2$& \code{truncpow4(x)} \\
Cauchy & $\theta_1(1 + \vert x_i - x_{i'}\vert^{\theta_2})^{-\frac{\theta_3}{\theta_2}}$ & $\theta_1, \theta_3 > 0$, $0 < \theta_2 \leq 2$ & \code{cauchy(x)} \\
Cauchy 3 & $\theta_1(1 + \vert x_i - x_{i'}\vert^{\theta_2})^{-3}$ & $\theta_1 > 0$, $0 < \theta_2 \leq 2$ & \code{cauchy3(x)} \\
\textit{Gaussian process} &&& \\
\textit{approximations} &&& \\
Nearest Neighbour & See section \ref{sec:gpapprox} & & \code{nngp_*(x)} \\
Hilbert Space & See section \ref{sec:gpapprox} & & \code{hsgp_*(x)} \\
\hline
\end{tabular}
\caption{\label{table:cov_fun}Supported covariance functions. $\vert . \vert$ is the Euclidean distance. $K_a$ is the modified Bessel function of the second kind. *Truncated power and Cauchy functions are defined for $\vert x_i - x_{i'} \vert < 1$ and are otherwise equal to zero and are used for sparse matrix generation, see Section \ref{subsec:compact}. }
\end{table}

One combines smaller functions to provide the desired overall covariance function. For example, for a stepped-wedge cluster randomised trial we could consider the standard specification (see, for example, \citet{Li2021}) with an exchangeable random effect for the cluster level (\code{j}), and a separate exchangeable random effect for the cluster-period (with \code{t} representing the discrete time period), which would be \code{~(1|gr(j))+(1|gr(j,t))}. Alternatively, we could consider an autoregressive cluster-level random effect that decays exponentially over time so that in a linear mixed model, for person $i$ in cluster $j$ at time $t$, $Cov(y_{ijt},y_{i'jt}) = \theta_1^2$, for $i\neq i'$, $Cov(y_{ijt},y_{i'jt'}) = \theta_1^2 \theta_2^{\vert t-t' \vert}$ for $t \neq t'$, and $Cov(y_{ijt},y_{i'j't}) = 0$ for $j \neq j'$. This function would be specified as \code{~(1|gr(j)*ar1(t))}. We could also supplement the autoregressive cluster-level effect with an individual-level random intercept, where \code{ind} specified the identifier of the individual, as \code{~(1|gr(j)*ar1(t)) + (1|gr(ind))}, and so forth. To add a further random parameter on some covariate \code{x} we must add an additional term, for example, \code{~(1|gr(j)*ar1(t)) + (1|gr(ind)) + (x|gr(ind))}. As another example, to implement a spatial Gaussian process model with exponential covariance function and two Cartesian coordinates \code{x} and \code{y} with autoregressive temporal decay we can specify \code{~(1|ar(t)*fexp0(x,y)}.

\subsubsection{Covariance function parameters}
The \code{covariance} argument of the call to \code{Model} is optional and if used, is typically used to specify the values of the parameters in the model. It can also receive a list, which can contain covariance function parameters, a random effects specification, and a particular data frame.  Where parameter values are provided, the elements of the vector correspond to each of the functions in the covariance formula in the order they are written. For example,
\begin{itemize}
    \item Formula: \code{~(1|gr(j))+(1|gr(j,t))}; parameters: \code{c(0.05,0.01)} describes the covariance function for $i\neq i'$
\begin{equation*}
Cov(y_{ijt},y_{i'j't'}) = 
\begin{cases}
0.05 + 0.01 & \text{if } j=j', t=t' \\
0.05 & \text{if } j=j', t\neq t' \\
0 & \text{otherwise}
\end{cases}
\end{equation*}

    \item Formula: \code{~(1|gr(j)*fexp0(t))}; parameters: \code{c(0.05,0.8)} describes the covariance function
$$
Cov(y_{ijt},y_{i'j't'}) = 
\begin{cases}
0.05* \text{exp}(-\frac{\vert t-t' \vert}{0.8}) & \text{if } j=j' \\
0 & \text{otherwise}
\end{cases}
$$
\end{itemize}

The Euclidean distance is used for all the functions. For example \code{~(1|fexp0(x,y))} will generate a covariance matrix where the covariance between two observations with positions $s=(x,y)$ and $s' = (x',y')$ is $Cov(s,s') = \text{exp}(- \vert \vert s-s' \vert \vert/ \theta)$ where $||.||$ is the Euclidean distance. For some of the covariance functions, we provide two parameterisations (for example, \code{fexp} and \code{fexp0}). The aim of these functions is to provide a version that is compatible with other functions with leading covariance parameters as, for example, a model specified as \code{gr(j)*fexp(t)} would not be identifiable as there would be two free parameters multiplying the exponential function.

\subsection{Covariance Calculation and Storage}
\label{subsec:covcalc}
The formula specifying the random effects is translated into a form that can be used to efficiently generate the value at any position in the matrix $D$ so that it can be quickly updated when the parameter values change, which facilitates functionality such as model fitting when using an arbitrary combination of covariance functions.

The matrices $D$ and $L$ are stored using sparse compressed row storage (CRS) format since for a large number of cases $D$ and its Cholesky factorisation are sparse. For example, a large number of covariance function specifications and their related study designs lead to a block diagonal structure for $D$:
\begin{equation}
\label{eq:blockd}
   D =  \begin{bmatrix}
    D_1 & \mathbf{0} & \hdots & \mathbf{0} \\
    \mathbf{0} &  D_2 & \hdots & \mathbf{0}\\
    \vdots & \vdots &\ddots & \vdots \\
    \mathbf{0} & \mathbf{0} & \hdots &  D_B 
    \end{bmatrix}
\end{equation}
Internally, the formula into parsed into blocks, both determined by the presence of a \code{gr} function in any particular random effect specification, and for each additive component of the specification. These blocks may be of differing dimensions and have different formulae and parameters. The data defining each element is a Euclidean distance and within each (likely dense) block, only the data for the strictly lower triangular portion needs to be stored. These distances are calculated once and stored in a flat data structure. We use a reverse Polish notation for to specify the instructions for a given block, as with the linear predictor, along with parameter indices and other necessary information. 

By default, the Cholesky decomposition $L$, such that $D = LL^T$, is calculated using a sparse matrix LDL algorithm \citep{Davis2005}, which also provides an efficient forward substitution algorithm for computing quadratic forms in the multivariate Gaussian likelihood (see Section \ref{sec:mcml}). Additionally, an efficient permutation is calculated on instantiation of the class using the approximation minimum degree algorithm \citep{Furrer2006,Amestoy2004}, which may provide a performance gain when factorising the matrix $PDP^T$ by ensuring the resulting factorisation is sparse. We provide a basic \proglang{C++} sparse matrix class, the LDL decomposition, and basic operations in the \pkg{SparseChol} package for \proglang{R}. 

As an alternative we can directly exploit the block diagonal structure of $D$. We can directly generate the decomposition of each block $L_b$ separately. We use the Cholesky–Banachiewicz algorithm for this purpose. Many calculations can also be broken down by block, for example $X^T\Sigma^{-1}X = \sum_b X^T_b L_b^{-T}L_b^{-1}X_b$. This example can also be further simplified using a forward substitution algorithm (see Section \ref{subsec:covparest}). To switch between sparse and dense matrix methods, one can use \code{model\$sparse(TRUE)} and \code{model\$sparse(FALSE)}, respectively. The default is to use sparse methods. 

\subsection{Compactly Supported Covariance Functions}
\label{subsec:compact}
For many types of GLMM, the matrix $D$ is dense and not blocked or sparse. For example, geospatial statistical models often specify spatial or spatio-temporal Gaussian process models for the random effects \citep{Diggle1998}. For even moderately sized data sets, this can result in very slow model fitting for the multivariate Gaussian likelihood. There are several approaches to reducing model fitting time. One approach is to use a compactly supported covariance function that leads to a sparse matrix $D$ while still approximating a Gaussian process. This method can be viewed as `covariance tapering' \citep{Kaufman2008}, although the approach implemented here is what they describe as the `one taper' method, which may be biased in some circumstances. As shown in Table \ref{table:cov_fun}, we include several compactly supported and parameterised covariance functions, which are described in \citet{Gneiting2002}. A covariance function with compact support has value zero beyond some `effective range'. Using these functions can result in a sparse matrix $D$, which is then amenable to the sparse matrix methods in this package. To implement these functions with an effective range of $r$, beyond which the covariance is zero, one must divide each variable by $r$. For example, if there are two spatial dimensions, \code{x} and \code{y}, in the data frame \code{data} then one would create \code{data\$xr <- data\$x/r}, and equivelently for \code{y}. Then, one could use the covariance function, for example, \code{~ truncpow3(xr,yr)}. 

\subsection{Gaussian Process Approximations}
\label{sec:gpapprox}
As an alternative to compactly supported covariance functions, one can instead use a Gaussian process approximation for large, dense covariance matrices. We provide two such approximations in this package.

\subsubsection{Nearest Neighbour Gaussian Process}
The multivariate Gaussian likelihood can be rewritten as the product of the conditional densities:
\begin{equation*}
    f(\mathbf{u}) = f(u_1) \prod_{i=2}^n f(u_i|u_1,...,u_{i-1})
\end{equation*}
\citet{Vecchia1988} proposed that one can approximate $f(\mathbf{u})$ by limiting the conditioning sets for each $u_j$ to a maximum size of $m$. For geospatial applications, \citep{Datta2016,Datta2016b} proposed the nearest neighbour Gaussian process (NNGP), in which the conditioning sets are limited to the $m$ `nearest neighbours' of each observation. 

Let $\mathcal{N}_j$ be the set of up to $m$ nearest neighbours of $j$ with index less than $j$. The approximation is:
\begin{equation*}
    f(\mathbf{u}) \approx f(u_1) \prod_{i=2}^n f(u_i|\mathcal{N}_i)
\end{equation*}
which leads to:
\begin{align}
\begin{split}
\label{eq:za}
    u_1 &= \eta_1 \\
    u_j &= \sum_{i=1}^p a_{ji}u_{\mathcal{N}_{ji}}
    \end{split}
\end{align}
where $\mathcal{N}_{ji}$ is the $i$th nearest neighbour of $j$. Equation (\ref{eq:za}) can be more compactly written as $\mathbf{u} = A\mathbf{u} + \boldsymbol{\eta}$ where $A$ is a sparse, strictly lower triangular matrix, $\boldsymbol{\eta} \sim N(0,\Omega)$ with $\Omega$ a diagonal matrix with entries $\Omega_{11} = \text{Var}(u_1)$ and $\Omega_{ii} = \text{Var}(u_i | \mathcal{N}_i)$ \citep{Finley2019}. The approximate covariance matrix can then be written as $D \approx (I-A)^{-1}\Omega(I-A)^{-T}$. We implement the algorithms described by \citet{Finley2019} to generate the matrices $A$ and $D$. The cost of generating these matrices is $O(nm^3)$ owing to the need to solve $n-1$ linear systems of up to size $m$. An efficient algorithm for calculating the quadratic form in the mulivariate Gaussian likelihood is also used.

A nearest neighbour approximation can be used by prefixing \code{nngp_} to the covariance function name with most of the functions in the package. For example, a nearest neighbour Gaussian process approximation using an exponential covariance function in two dimensions can be used by specifying in the formula \code{(1|nngp_fexp(x,y))}. The approximation parameters including the number of nearest neighbours can be set (or returned) using the function \code{model$covariance$nngp()}.

\subsubsection{Hilbert Space Gaussian Process}
Low, or reduced, rank approximations aim to approximate the matrix $D$ with a matrix $\Tilde{D}$ with rank $m < n$. The optimal low-rank approximation is $\Tilde{D} = \Phi \Lambda \Phi^T$ where $\Lambda$ is a diagonal matrix of the $m$ leading eigenvalues of $D$ and $\Phi$ the matrix of the corresponding eigenvectors. However, the computational complexity of generating the eigendecomposition scales the same as matrix inversion. \citet{Solin2020} propose an efficient method to approximate the eigenvalues and eigenvectors using Hilbert space methods, so we refer to it as a Hilbert Space Gaussian Process (HSGP). \citet{RiutortMayol2023} provides further discussion of these methods.

Stationary covariance functions, including those in the Matern class like exponential and squared exponential, can be represented in terms of their spectral densities. For example, the spectral density function of the squared exponential function in $D$ dimensions is:
\begin{equation*}
S(\omega) = \sigma^2 (\sqrt{2\pi})^D \phi^D \text{exp}(-\phi^2 \omega^2/2)
\end{equation*}

Consider first a unidimensional space with support on $[-c,c]$. The eigenvalues $\lambda_j$ (which are the diagonal elements of $\Lambda$) and eigenvectors $\phi_j$ (which form the columns of $\Phi$) of the Laplacian operator in this domain are:
\begin{equation*}
\lambda_j = \left( \frac{j\pi}{2L} \right)^2 
\end{equation*}
and
\begin{equation*}
\phi_j(x) = \sqrt{\frac{1}{L}} \text{sin}\left(  \sqrt{\lambda_j}(x+L) \right)
\end{equation*}
Then the approximation in one dimension is
\begin{equation*}
\mathcal{Z}(x) \approx \sum_{j=1}^m S\left(\sqrt{\lambda_j}\right)^{1/2}\phi_j(x)\beta_j
\end{equation*}
where $\beta_j \sim N(0,1)$. This result can be generalised to multiple dimensions. The total number of eigenvalues and eigenfunctions in multiple dimensions is the combination of all univariate eigenvalues and eigenfunctions over all dimensions. The matrix $\Phi$ does not depend on the covariance parameters are can be pre-computed, so only the product $\Phi\Lambda^{\frac{1}{2}}$ needs to be re-calculated during model fitting, which scales as $O(nm^2)$. \citet{RiutortMayol2023} provide a detailed analysis of the reduced rank GP for unidimensional linear Gaussian models, examining in particular how the choice of $c$ and $p$ affect performance of posterior inferences and model predictions. 

The HSGP is only currently available with the exponential or squared expoential covariance functions as either \code{hsgp_fexp(x)} or \code{hsgp_sqexp(x)}. The approximation parameters including the number of basis functions and the boundary condition can be set (or returned) using the function \code{model$covariance$hsgp()}.

\subsection{Computation of matrix Z}
\label{sec:matZ}
The matrix $Z$ is constructed in a similar way as other packages, such as described for \pkg{lme4} \citep{lme4}. $Z$ is comprised of the matrices corresponding to each block $Z = [Z_1, Z_2, ..., Z_B]$. For a model formula \code{(1|f(x1,x2,...))}, the dimension of the corresponding matrix $Z_b$ is $n$ rows and number of columns equal to the number of unique rows of the combination of the variables in \code{x1,x2,...}. The $i$th row and $j$th column of $Z_b$ is then equal to 1 if the $i$th individual has the value corresponding to the $j$th unique combination of the variables and zero otherwise. For formulae specifying random effects on covariates (``variable slopes models''), e.g. \code{(z|f(x1,x2,...))}, then the $i$th row and $j$th column of $Z_b$ is further multiplied by $x_i$. $Z$ is also stored in CRS format and uses sparse matrix methods for multiplication and related calculations.

\subsection{Additional Arguments}
\label{sec:addarguments}
For Gaussian models, and other distributions requiring an additional scale parameter $\phi$, one can also specify the option \code{var_par} which is the conditional variance $\phi = \sigma$ at the individual level. The default value is 1. Currently (version 0.6.1), the package supports the following families (link functions): Gaussian (identity, log), Poisson (log, identity), Binomial (logit, log, probit, identity), Gamma (log, inverse, identity), and Beta (logit). For the beta distribution one must provide \code{Beta()} to the \code{family} argument.

\subsection{Approximation of the Covariance Matrix}
\label{sec:approxsigma}
The \code{Model} class can provide an approximation to the covariance matrix $\Sigma$ with the member function \code{\$Sigma()}. This approximation is also used to calculate the expected information matrix with \code{\$information\_matrix()} and estimate power in the member function \code{\$power()} (see Section \ref{subsec:power}). We use the first-order approximation based on the marginal quasi-likelihood proposed by \citet{breslow1993approximate}:
\begin{equation}
    \Sigma = W^{-1} + ZDZ^T
\end{equation}
where $W^{-1} = \text{diag}\left( \left(\frac{\partial h^{-1}(\boldsymbol{\eta})}{\partial \boldsymbol{\eta}}\right)^2 \text{Var}(\mathbf{y}| \mathbf{u})\right)$, which are recognisable as the GLM iterated weights \citep{breslow1993approximate, mccullagh2019generalized}. For Gaussian-identity mixed models this approximation is exact. The diagonal of the matrix $W$ can be obtained using the member function \code{\$w\_matrix()}. The information matrix for $\beta$ is:
\begin{equation}
\label{eq:infomat}
    M_\beta = (X^T\Sigma^{-1}X)^{-1}
\end{equation}

\citet{Zeger1988} suggest that when using the marginal quasilikelihood, one can improve the approximation to the marginal mean by ``attenuating'' the linear predictor for non-linear models. For example, with the log link the ``attenuated'' mean is $E(y_i) \approx h^{-1}(x_i\beta + z_iDz_i^T/2)$ and for the logit link  $E(y_i) \approx h^{-1}(x_i\beta\text{det}(a Dz_i^Tz_i + I)^{-1/2})$ with $a = 16\sqrt{3}/(15\pi)$. To use ``attenuation'' one can set \code{mod\$use\_attenutation(TRUE)}, the default is not to use attenutation.

\subsection{Updating Parameters}
\label{subsec:updating}
The parameters of the covariance function and linear predictor can updated using the function \code{update\_parameters()}, which then triggers re-generation of the matrices and updates the data in the underlying \proglang{C++} classes: \code{model\$update_parameters( cov.pars = c(0.1,0.9))}.

\subsection{Class Inheritance}
The \pkg{R6} class system provides many of the standard features of other object orientated systems, including class inheritance. The classes specified in the \pkg{glmmrBase} package can be inherited from to provide the full range of calculations to other applications. As an example we can define a new class that has a member function that returns the log determinant of the matrix $D$:
\begin{CodeChunk}
\begin{CodeInput}
CovDet <- R6::R6Class("CovDet",
                       inherit = Covariance,
                       public = list(
                       det = function(){
                         return(Matrix::determinant(self$D)$modulus)
                       }))
cov <- CovDet$new(formula = ~(1|gr(j)*ar1(t)),
                      parameters = c(0.05,0.8),
                      data= data)
cov$det()
[1] -72.26107
\end{CodeInput}
\end{CodeChunk}
More complex applications may include classes to implement simulation-based analyses that simulate new data and fit a GLMM using one of the package's model fitting methods.

\subsection{Power Calculation}
\label{subsec:power}
Power and sample size calculations are an important component of the design of many studies, particularly randomised trials, which we use as an example. Cluster randomised trials frequently use GLMMs for data analysis, and hence are the basis for estimating the statistical power of a trial design \citep{Hooper2016}. Different approaches are used to estimate power or calculate sample size given an assumed correlation structure, most frequently using ``design effects'' \citep{Hemming2020}. However, given the large range of possible models and covariance structures, many software packages implement a narrow range of specific models and provide wrappers to other functions. For example, we identified eight different \proglang{R} packages available on CRAN or via a R Shiny web interface for calculating cluster trial power or sample size. These packages and their features are listed in Tables \ref{table:features} and \ref{table:supported_models}. As is evident, the range of models is relatively limited. Beyond this specific study type the R package \pkg{simr} provides functions to simulate data from GLMMs and estimate statistics like power using Monte Carlo simulation for specific designs. 

\begin{table}
\centering
\small
\begin{tabular}[t]{l|l|l|l|l|l|l}
\hline
Package & \shortstack{Custom \\ Designs} & \shortstack{Data \\simulation} & \shortstack{Power by\\ simulation} & \shortstack{Power by \\approx.} & \shortstack{Non-simple\\ randomisation} &  \shortstack{Optimal\\ designs}\\
\hline
\pkg{SWSamp} & \checkmark & \checkmark & \checkmark & \checkmark & $\times$ & $\times$\\
\hline
\pkg{samplingDataCRT} & \checkmark & \checkmark & $\times$ & \checkmark & $\times$ &  $\times$\\
\hline
\pkg{ShinycRCT} & \checkmark & $\times$ & $\times$ & \checkmark & $\times$ &  $\times$\\
\hline
\pkg{swCRTdesign} & \checkmark & $\times$ & $\times$ & \checkmark & $\times$ & $\times$\\
\hline
\pkg{clusterPower} & $\sim^1$ & $\times$ & \checkmark & \checkmark & $\sim^3$ & $\times$\\
\hline
\pkg{CRTdistpower} & \checkmark & $\times$ & $\times$ & \checkmark & $\times$  & $\times$\\
\hline
\pkg{swdpwr} & \checkmark & $\times$ & $\times$ & \checkmark & $\times$ &  $\times$\\
\hline
\pkg{SteppedPower} & \checkmark & $\times$ & $\times$ & \checkmark & $\times$ & $\times$\\
\hline
\pkg{glmmrBase} & \checkmark & \checkmark & \checkmark & \checkmark &  \checkmark & \checkmark\\
\hline
\end{tabular}
\caption{\label{table:features}Features of packages for \proglang{R} to calculate power and sample size for cluster randomised trial designs}
\end{table}

\begin{table}
\centering
\scriptsize
\begin{tabular}[t]{l|l|l|l|l|l|l|l|l}
\hline
Package & \shortstack{Non-\\canonical\\ link} & \shortstack{Binom./\\ Poisson} & \shortstack{Other\\dist.} & \shortstack{Compound\\symmetry} & \shortstack{Temporal\\ decay} & \shortstack{Random\\slopes} & Covariates & \shortstack{Other\\ functions}\\
\hline
\pkg{SWSamp} & $\times$ & \checkmark & $\times$ & \checkmark & $\times$ & $\times$ & $\times$ & $\times$\\
\hline
\pkg{samplingDataCRT} & $\times$ & $\sim$ & $\times$ & \checkmark & $\times$ & $\times$ & $\times$ & $\times$\\
\hline
\pkg{ShinycRCT} & $\times$ & \checkmark & $\times$ & \checkmark & \checkmark & $\times$ & $\times$ & $\times$\\
\hline
\pkg{swCRTdesign} & $\times$ & \checkmark & $\times$ & \checkmark & $\times$ & $\times$ & $\times$ & $\times$\\
\hline
\pkg{clusterPower} & $\times$ & \checkmark & $\times$ & \checkmark & $\times$ & $\times$ & $\times$ & $\times$\\
\hline
\pkg{CRTdistpower} & $\times$ & \checkmark & $\times$ & $\sim^2$ & $\times$ & $\times$ & $\times$ & $\times$\\
\hline
\pkg{swdpwr} & \checkmark & \checkmark & $\times$ & \checkmark & $\times$ & $\times$ & $\times$ & $\times$\\
\hline
\pkg{SteppedPower} & $\times$ & $\sim$ & $\times$ & \checkmark & $\times$ & \checkmark & $\sim^3$ & $\times$\\
\hline
\pkg{glmmrBase} & \checkmark & \checkmark & \checkmark & \checkmark & \checkmark & \checkmark & \checkmark & \checkmark\\
\hline
\end{tabular}
\caption{\label{table:supported_models}Supported models of packages for \proglang{R} to calculate power and sample size for cluster randomised trial designs}
\end{table}

As an alternative, the flexible model specification and range of functionality of \pkg{glmmrBase} can be used to quickly estimate the power for a particular design and model. The \code{Model} class member function \code{\$power()} estimates power by calculating the information matrix $M$ described above. The square root of the diagonal of this matrix provides the (approximate) standard errors of $\beta$. The power of a two-sided test for the $i$th parameter at a type I error level of $\alpha$ is given by $\Phi(\vert\beta_i\vert/\sqrt{M_{ii}} - \Phi^{-1}(1-\alpha/2))$ where $\Phi$ is the Gaussian cumulative distribution function. As an example, to estimate power for a stepped-wedge parallel cluster randomised trial (see \citet{Hooper2016}) with 10 clusters and 11 time periods and ten individuals per cluster period, where we use a Binomial-logit model with time period fixed effects set to zero, and use a auto-regressive covariance function with parameters 0.25 and 0.7:
\begin{CodeChunk}
\begin{CodeInput}
data <- nelder(~(cl(10) * t(11)) > i(10))
data$int <- 0 # int is the intervention effect
data[data$t > data$cl,'int'] <- 1
model <- Model$new(formula = ~ factor(t) + int - 1 + (1|gr(cl)*ar1(t)),
                   data = data,
                   covariance = c(0.05,0.7),
                   mean = c(rep(0,12),0.5),
                   family = binomial())
model$power()
\end{CodeInput}
\begin{CodeOutput}
    Parameter   Value  SE        Power
...
12         int   0.5 0.1816136 0.7861501
\end{CodeOutput}
\end{CodeChunk}
One can use the functionality of the \code{Model} class to investigate power for a range of model parameters. For example, to estimate power for covariance parameters in the ranges (0.05,0.5) and (0.2,0.9) we can create a data frame holding the relevant combination of values and iteratively update the model:
\begin{CodeChunk}
\begin{CodeInput}
param_data <- expand.grid(par1 = seq(0.05,0.5,by=0.05), 
                          par2 = c(0.2,0.9,by=0.1),power=NA)
for(i in 1:nrow(param_data)){
    model$update_parameters(cov.pars = 
            c(param_data$par1[i], param_data$par2[i]))
    param_data$power[i] <- model$power()[12,3]
}
head(param_data)
\end{CodeInput}
\begin{CodeOutput}
  par1 par2     power
1 0.05  0.2 0.8348863
2 0.10  0.2 0.7852298
3 0.15  0.2 0.7381658
4 0.20  0.2 0.6945137
5 0.25  0.2 0.6544970
6 0.30  0.2 0.6180314
\end{CodeOutput}
\end{CodeChunk}

\subsection{Data Simulation}
\label{sec:datasim}
The \code{glmmrBase} package also simplifies data simulation for GLMMs. Data simulation is a commonly used approach to evaluate the properties of estimators and statistical models. The \code{Model} class has member function \code{sim\_data()}, which will generate either vector of simulated output $y$ (argument \code{type="y"}), a data frame that combines the linked dataset with the newly simulated data (argument \code{type="data"}), or a list containing the simulated data, random effects, and matrices $Z$ and $X$ (argument \code{type="all"}). To quickly simulate data new random effects are drawn as $\mathbf{v} \sim N(0,1)$ and then the random effects component of the model simulated as $ZL\mathbf{v}$. Simulation only of the random effects terms can be done with the relevant function in the \code{Covariance} class, \code{\$covariance\$simulate\_re()}.

\section{Model Fitting}
\label{sec:mcml}
The parameters and standard errors of GLMMs are difficult to estimate in a maximum likelihood context given the intergral in the likelihood in Equation \ref{eq:lik1}. Several approximations and associated software are available for GLMMs, however, there are few options for fitting more complex models. The \code{Model} class includes two functions: \code{MCML}, for MCML model fitting, and \code{LA}, which uses a Laplace approximation, both described below. 

\subsection{Existing software}
\label{subsec:software}
Software for fitting generalised linear mixed models has been available in \proglang{R} environment \citep{r_lang} for many years. The widely-used package \pkg{lme4} provides maximum likelihood and restricted maximum likelihood (REML) estimators \citep{lme4}. \pkg{lme4} builds on similar methods proposed by e.g. \citet{Bates2004} and \citet{Henderson1982}. For non-linear models it uses Laplacian approximation or adaptive Gaussian quadrature. However, \pkg{lme4} only allows for compound symmetric (or exchangable) covariance structures, i.e. group-membership type random effects, and linear, additive linear predictors. Function \code{glmmPQL()} in package \pkg{MASS} \citep{mass} implements Penalized quasi-likelihood (PQL) methods for GLMMs, but is again limited to the same covariance structures, and linear, additive model structure. The package \pkg{nlme} \citep{nlme} also uses REML approaches for linear mixed models and approximations for models with mean specifications with non-linear functions of covariates discussed in \cite{lindstrom1990nonlinear}. \pkg{nlme} provides a wider range of covariance structures, such as autoregressive functions, which are also available nested within a broader group structure for linear, additive linear predictors. The package providing Frequentist model estimation of GLMMs with the broadest support for different specifications, including many different covariance functions, is \pkg{glmmTMB}, which uses the \pkg{TMB} (Template Model Builder) package to support model specification and implementation of Laplacian Approximation to estimate models, although also only permits linear, additive linear predictors. 

Outside of R, GLMM model fitting is also provided in other popular statistical software packages. Stata offers a range of functions for fitting mixed models. The \code{xtmixed} function offers both maximum likelihood and restricted maximum likelihood methods to estimate linear mixed models with multi-level group membership random effect structures and heteroscedasticity. For GLMMs, there is a wide range of models that allow for random effects, many of which are accessed through the \code{meglm} suite of commands. Estimation can use Gauss-Hermite quadrature or Laplace approximation, but random effect structures are generally limited to group membership type effects. For the software SAS, \code{PROC MIXED}, and \code{PROC GLIMMIX} provide GLMM model fitting using either quadrature or Laplace approximation methods. These commands provide a relatively broad random of random effect coviariance structures, including autoregressive, Toeplitz, and exponential structures. \code{PROC NLMIXED} allows for non-linear functions of fixed and random effects; it uses a Gaussian quadrature method to approximate the integral in the log-likelihood, and a Newton-Raphson step for empirical Bayes estimation of the random effects. However, \code{PROC NLMIXED} only allows for group membership random effects. Both Stata and SAS also provide a range of options for standard error estimation and correction. Both software packages are proprietary.

We also provide in-built robust and bias-corrected standard error estimates where required. In R, the packages \pkg{lmerTest} and \pkg{pbkrtest} provide a range of corrections to standard errors for models fit with \code{lme4}. The package architecture means that \pkg{glmmrBase} can provide these standard errors for the range of models available in the package, and generally in less time as they do not require refitting of the model. Stata and SAS can calculate bias-corrected and robust standard error estimates with the commands described above. 

We also note that there are also several packages for Bayesian model fitting in R, including \pkg{brms} \citep{Burkner2017} and \pkg{rstanarm}, which interface with the Stan statistical software (described below), and \pkg{MCMCglmm}. Stan is a fully-fledged probabilistic programming language that implements MCMC, and allows for a completely flexible approach to model specification \citep{carpenter2017stan}. Stan can also provide limited maximum likelihood fitting functionalist and related tools for calculating gradients and other model operations.

Evidently, there already exists a good range of support for GLMM model fitting available in \proglang{R}. \pkg{glmmrBase} may be seen to provide a complement to existing resources rather than another substitute. As we discuss below, in a maximum likelihood context, the existing software all provide approximations rather than full likelihood model fitting. These approximations are typically accurate and fast, but they can also frequently fail to converge and in more complex models the quality of the approximation may degrade, which has lead to methods for bias corrections and improvements to these approximations (e.g. \citet{Breslow1995,Lin1996,Shun1995,Capanu2013}). \pkg{glmmrBase} provides an alternative that can be used to support more complex analyses, flexible covariance specification, and provide a comparison to ensure approximations are reliable, while also providing an array of other functionality.

\subsection{Stochastic Maximum Likelihood Estimation}
\label{subsec:mcml}
Markov Chain Monte Carlo Maximum Likelihood (MCML) are a family of algorithms for estimating GLMMs using the full likelihood that treat the random effect terms as missing data, which are simulated on each iteration of the algorithm. Estimation can then be achieved to an arbitrary level of precision depending on the number of samples used. Approximate methods such as Laplace approximation or quadrature provide significant computational advantages over such approaches, however they may trade-off the quality of parameter and variance estimates versus full likelihood approaches. Indeed, the quality of the Laplace approximation method for GLMMs can deteriorate if there are smaller numbers of clustering groups, smaller numbers of observations to estimate the variance components, or if the random effect variance is relatively large.

The package provides full likelihood model fitting for the range of GLMMs that can be specified using the package via the MCML algorithms described by \cite{mcculloch1997maximum} or stochastic approximation expectation maximisation (SAEM) \citep{Jank2006}. Each algorithm has three main steps per iteration: draw samples of the random effects using Markov Chain Monte Carlo (MCMC), the mean function parameters are then estimated conditional on the simulated random effects, and then the parameters of the multivariate Gaussian likelihood of the random effects are then estimated. The process is repeated until convergence. We describe each step in turn and optimisations.

\subsection{MCMC sampling of random effects}
\label{sec:mcmc}
Reliable convergence of the algorithm requires a set of $m_i$ independent samples of the random effects on the $i$th iteration, i.e. realisations from the distribution of the random effect conditional on the observed data, and with fixed parameters. MCMC methods can be used to sample from the `posterior density' $f_{\mathbf{u}|\mathbf{y}}(\mathbf{u}|\mathbf{y},\beta^{(i)},\phi^{(i)},\theta^{(i)}) \propto f_{\mathbf{y}|\mathbf{u}}(\mathbf{y}|\mathbf{u},\beta^{(i)},\phi^{(i)})f_{\mathbf{u}}(\mathbf{u}|\theta^{(i)})$. To improve sampling, instead of generating samples of $\mathbf{u}$ directly, we instead sample $\mathbf{v}$ from the model:
\begin{align}
\begin{split}
\label{eq:mcmc}
    \mathbf{y} &\sim G(h(g(X,\hat{\beta}) + \Tilde{Z}\mathbf{v}); \hat{\phi}) \\
    \mathbf{v} &\sim N(0,I)
    \end{split}
\end{align}
where $\Tilde{Z} = ZL$, and $I$ is the identity matrix. Once $m_i$ samples of $\mathbf{v}$ are returned, we then transform the samples as $\mathbf{u} = L\mathbf{v}$. 



We use Stan for MCMC sampling. Stan implements Hamiltonian Monte Carlo (HMC) with advances such as the No-U-Turn Sampler (NUTS) \citep{carpenter2017stan,Homan2014}, which automatically `tunes` the HMC parameters. Stan can generally achieve a good effective sample size per unit time and so lead to better convergence of the MCMCML algorithm than standard MCMC algorithms.

Each iteration of the MCMC sampling must evaluate the log-likelihood (and its gradient). Since the elements of both $\mathbf{y}$ and $\mathbf{v}$ are all independent in this version of the model, evaluating the log-likelihood is parallelisable within a single chain, which significantly improves running time.

\subsection{Estimation of mean function parameters}
\label{sec:estbeta}
There are several methods to generate new estimates of $\beta$ and $\phi$ conditional on $\mathbf{y}$, $\mathbf{u}^{(i)}$, and $\phi^{(i)}$. We aim to minimise the negative log-likelihood:
\begin{equation}
\label{eq:betalogl}
    \mathcal{L}(\beta|\mathbf{u}) = E_u\left[ - \log f_{\mathbf{y}|\mathbf{u}}(\mathbf{y}|\mathbf{u},\beta,\phi) \right] \approx \frac{1}{m_i}\sum_{k=1}^{m_i} - \log f_{\mathbf{y}|\mathbf{u}}(\mathbf{y}|\mathbf{u}^{(i,k)},\beta,\hat{\phi})
\end{equation}
There are three different approaches:

\begin{itemize}
    \item \textit{Monte Carlo Newton Raphson (MCNR)}: If the linear predictor is linear and additive in the data and parameters, then we can use the explicit formulation:
\begin{equation*}
        \beta^{(i+1)} = \beta^{(i)} + E_u \left[ X^T W(\theta^{(i)},\mathbf{u}) X \right] ^{-1}X^T \left( E_u \left[ W(\theta^{(i)},\mathbf{u}) \frac{\partial h^{-1}(\boldsymbol{\eta})}{\partial \boldsymbol{\eta}} (\mathbf{y - \boldsymbol{\mu}}(\beta^{(i)},\mathbf{u}))|\mathbf{y} \right] \right)
\end{equation*}
where the expectations are with respect to the random effects. We estimate the expectations using the realised samples from the MCMC step of the algorithm. For many GLMMs, the likelihoods can be multimodal \citet{Searle1992}, which means the MCNR algorithm may fail to converge at all.
\item \textit{Monte Carlo Quasi-Newton (MCQN)}: As the gradient is relatively inexpensive to calculate, one can make use of quasi-Newton algorithms, including L-BFGS. 
\item \textit{Monte Carlo Expectation Maximisation (MCEM)}: We use the BOBYQA algorithm, a numerical, derivative free optimser that allows bounds for expectation maximisation \citep{powell2009bobyqa}. 
\end{itemize}

\subsection{Estimation of covariance parameters}
\label{subsec:covparest}
The final step of each iteration is to generate new estimates of $\theta$ given the samples of the random effects. We aim to minimise 
\begin{equation}
\label{eq:thetalogl}
    \mathcal{L}(\theta | \mathbf{u}) = E_{\mathbf{u}} \left[ -\log(f_{\mathbf{u}}(\mathbf{u}|\theta)) \right] \approx \frac{1}{m_i}\sum_{k=1}^{m_i} \log(f_{\mathbf{u}}(\mathbf{u}^{(i,k)}|\theta))
\end{equation}
The multivariate Gaussian density is:
\begin{align*}
    \log f_{\mathbf{u}}(\mathbf{u}|\theta) &= -\frac{m}{2}\log(2\pi) - \frac{1}{2}\log(|D|) - \frac{1}{2}\mathbf{u}^T D^{-1} \mathbf{u} \\
    &= -\frac{m}{2}\log(2\pi) - \frac{1}{2}\sum_{j = 1}^{n} \log(L_{jj}) - \frac{1}{2}\mathbf{u}^T L^{-T}L^{-1} \mathbf{u}
\end{align*}
where $L_{jj}$ is the $j$th diagonal element of $L$.

As described in Section \ref{subsec:compact}, we can take two approaches to the Cholesky factorisation of matrix $D$, either exploiting the block diagonal structure or using sparse matrix methods. In both cases the quadratic form can be rewritten as $\mathbf{u}^T D^{-1} \mathbf{u} = \mathbf{u}^T L^{-T}L^{-1} \mathbf{u} = \mathbf{z}^T\mathbf{z}$, so we can obtain $\mathbf{z}$ by solving the linear system $L\mathbf{z} = \mathbf{u}$ using forward substitution, which, if $L$ is also sparse provides efficiency gains. The log determinant is $\log(|D|) = \sum \log(\text{diag}(L))$. In the case of using a block diagonal formulation, we calculate the form:
\begin{equation*}
    \log f_{\mathbf{u}}(\mathbf{u}|\theta) = \sum_{b=0}^{\texttt{B}} \left(-\frac{m}{2}\log(2\pi) - \frac{1}{2}\sum_{i = r_b}^{s_b} \log(\text{diag}(L)_i) - \frac{1}{2}\mathbf{z}^T_{r_b:s_b} \mathbf{z}_{r_b:s_b} \right)
\end{equation*}
where $r_b$ and $s_b$ indicates the start and end indexes of the vector, respectively, corresponding to block $b$. 

The gradient of the log-likelihood with respect to $\theta$ is:
\begin{align*}
    \nabla_\theta \log f_{\mathbf{u}}(\mathbf{u}|\theta) &= -\frac{1}{2}\text{trace}\left( D^{-1} \frac{\partial D}{\partial \theta} \right) - \frac{1}{2}\mathbf{u}^T D^{-1} \frac{\partial D}{\partial \theta} D^{-1} \mathbf{u} \\
    &= -\frac{1}{2} \text{trace}\left( L^{-T} L^{-1} \frac{\partial D}{\partial \theta} \right)  - \frac{1}{2}\mathbf{u}^T L^{-T} L^{-1} \frac{\partial D}{\partial \theta} L^{-T} L^{-1} \mathbf{u}
\end{align*}
The quadratic form can similarly be solved by applying forward and backward substitution. The L-BFGS-B algorithm can be used then to minimise the log-likelihood. However, the inversion of $D$ and calculation of its partial derivatives may make the gradient computationally demanding. A derivative-free algorithm, BOBYQA, while requiring more function evaluations may be more efficient at solving for the parameter values. Both methods are available.


\subsection{Stochastic Approximation Expectation Maximisation}
The efficiency of the MCML algorithms can be improved in two ways: first, one can dynamically alter the number of MCMC samples per step, which is discussed in the next section, and second, one can make use of MCMC sample information from previous iterations, reducing the number of samples per iteration. Stochastic Approximation Expectation Maximisation (SAEM) is a Ruppert-Monroe algorithm that can be used to approximate the log likelihoods (\ref{eq:betalogl}) and (\ref{eq:thetalogl}) \citep{Jank2006}. For $\beta$ the log-likelihood on the $i+1$th iteration is:
\begin{equation*}
    \hat{\mathcal{L}}^{(i+1)} = \hat{\mathcal{L}}^{(i)} + \gamma \left( 
 \frac{1}{m_i}\sum_{k=1}^{m_i} - \log f_{\mathbf{y}|\mathbf{u}}(\mathbf{y}|\mathbf{u}^{(i,k)},\beta,\hat{\phi}) - \hat{\mathcal{L}}^{(i)}\right)
\end{equation*}
and equivalently for $\theta$, where $\gamma \propto \left(\frac{1}{i}\right)^\alpha$ for $\alpha \in [0.5,1)$. This approach makes use of the all the MCMC samples from every iteration, where the contribution of previous values diminish over the iterations depending on the parameter $\alpha$. Larger values of $\alpha$ result in faster `forgetting'. For some applications, Ruppert-Polyak averaging can further improve convergence of the algorithms \citep{Polyak1992}, where the estimate of the log-likelihood on iteration $i$ is:
\begin{equation*}
    \Bar{\hat{\mathcal{L}}}^{(i)} = \frac{1}{i}\sum_{j=1}^i \hat{\mathcal{L}}^{(i)}
\end{equation*}

\subsection{Stopping Criteria and MCMC Sample Sizes}
We provide two types of rule for terminating the algorithm. First, one can monitor the differences in the parameter estimates between successive iterations, and terminate the algorithm when the largest difference falls below some tolerance. However, this method may lead to premature termination. A running mean could prevent such issues, however, a perhaps preferable method is to assess the changes in the estimates of the log-likelihood:
\begin{equation*}
    \Delta \hat{\mathcal{L}}^{(i)} = \hat{\mathcal{L}}^{(i+1)} - \hat{\mathcal{L}}^{(i)}
\end{equation*}
An expectation maximisation algorithm is guaranteed to improve the objective function value over time. However, the estimates of the log-likelihood are subject to Monte Carlo error and so $\Delta \hat{\mathcal{L}}^{(i)}$ may increase or decrease at convergence. \citet{Caffo2005} proposed a stopping criteria that monitors the probability of convergence. The estimated upper bound for the difference is:
\begin{equation*}
    U \hat{\mathcal{L}}^{(i)} = \Delta \hat{\mathcal{L}}^{(i)} + z_\gamma \hat{\sigma}_\Delta
\end{equation*}
where $z_\gamma$ is the $\gamma$ quantile of a standard Gaussian distribution and $\hat{\sigma}_\Delta$ is the estimated standard deviation of $\Delta \hat{\mathcal{L}}^{(i)}$. This upper bound will be negative with probability $\gamma$ at convergence. 

\citet{Caffo2005} use a similar argument to dynamically alter the number of MCMC samples on each iteration to acheive convergence with probability $\gamma$. The propose:
\begin{equation*}
    m_{i+1} = \max \left( m_i, (\mathcal{L}(\beta) + \mathcal{L}(\theta)\frac{z_\gamma + z_\epsilon}{\Delta \hat{\mathcal{L}}^{(i)}}  \right)
\end{equation*}
We provide MCML algorithm with and without dynamic sample sizes.

\subsection{Model Estimation}
The member function \code{MCML} fits the model. There are multiple options to control the algorithm and set its parameters; most options are determined by the relevant parameters of the \code{Model} object. The default fitting method is SAEM without Ruppert-Polyak averaging, and with the log-likelihood stopping rule with probability 0.95, as this has been found to provide the most efficient and reliable model fitting in many tests.

\subsection{Laplace Approximation}
\label{sec:LA}
We also provide model fitting using a Laplace approximation to the likelihood with the \code{Model} member function \code{LA()}. We include this approximation as it often provides a faster means of fitting the models in this package. These estimates may be of interest in their own right, but can also serve as useful starting values for the MCML algorithm. We approximate the log-likelihood of the re-parameterised model given in Equation (\ref{eq:mcmc}) as:
\begin{equation}
\label{eq:la1}
    \log L(\beta,\phi,\theta,\mathbf{v}|\mathbf{y}) \approx -\frac{1}{2}\vert I + L^TZ^T W ZL \vert + \sum_{i=1}^n\log(f_{y\vert\mathbf{u}}(y_i\vert \mathbf{v},\beta,\phi))) -\frac{1}{2}\mathbf{v}^T\mathbf{v}
\end{equation}
As before, we use the first-order approximation (with or without attenutation), $W^{-1}(\theta,\mathbf{u}) = \text{diag}\left( \left(\frac{\partial h^{-1}(\boldsymbol{\eta})}{\partial \boldsymbol{\eta}}\right)^2 \text{Var}(\mathbf{y}| \mathbf{u})\right)$. We iterate model fitting to obtain estimates of  $\beta$, $\mathbf{v}$ and $\phi$ and then to then obtain estimates of $\theta$ until the values converge. We provide two iterative algorithms for model fitting, using a similar pattern to the MCNR, and MCEM algorithms. First, the values of $\beta$ and $\mathbf{v}$ are updated. We can either use a scoring algorithm approach, or numerical optimisation. For the former method the score system we use is:
\begin{equation}
\label{eq:scoring}
    \begin{bmatrix}
        X^TWX & X^TWZL \\
        Z^TL^TWX & Z^TL^TWLZ + I
    \end{bmatrix} 
    \begin{bmatrix}
        \beta \\
        \mathbf{v}
    \end{bmatrix} = \begin{bmatrix}
        X^TW \frac{\partial \boldsymbol{\eta}}{\partial \boldsymbol{\mu}} (\mathbf{y} - \boldsymbol{\mu}) \\
        L^TZ^TW \frac{\partial \boldsymbol{\eta}}{\partial \boldsymbol{\mu}} (\mathbf{y} - \boldsymbol{\mu})
    \end{bmatrix}
\end{equation}
where $\frac{\partial \boldsymbol{\eta}}{\partial \boldsymbol{\mu}}$ is a diagonal matrix with elements $\frac{\partial \eta_i}{\partial \mu_i}$. Where there are non-linear functions of parameters in the linear predictor, a first-order approximation is used where $X$ is replaced by an $n \times P$ matrix with row $i$ equal to $J_\beta$ evaluated at the values of $x_i$. This system of score equations is highly similar to those used in other software packages and articles on approximate inference for GLMMs (e.g. \citet{Bates2004,lme4,breslow1993approximate}). The alternative approach fits $\beta$ and $\mathbf{v}$ by minimising the log-likelihood in Equation (\ref{eq:la1}) using the BOBYQA algorithm. In the second step, the variance parameters $\theta$ are updated by again minimising (\ref{eq:la1}), fixing $\beta$ and $\mathbf{v}$ to their current estimates.

\subsection{Standard errors}
\label{subsec:se}
There are several options for the standard errors of $\hat{\beta}$ and $\hat{\theta}$ returned by either \code{LA} or \code{MCML} and are accessible through a \code{Model} object. The currently implemented methods are described here.

\subsubsection{GLS standard errors}
The default GLS standard errors are given by the square root of the diagonal of $(X^T \Sigma^{-1} X)^{-1}$. The standard errors for the covariance parameter estimates are estimated from the inverse expected information matrix for $\theta$, $M_\theta$, the elements of which are given by:
\begin{equation*}
    [M_{\theta}]_{i,j} = \text{tr}\left( \frac{\partial \Sigma^{-1}}{\partial \theta_i} \Sigma \frac{\partial \Sigma^{-1}}{\partial \theta_j} \Sigma \right)
\end{equation*}
Using the expression for $\Sigma$ given in Section \ref{sec:approxsigma}, we can write 
\begin{equation*}
    A_i = \frac{\partial \Sigma}{\partial \theta_i} = Z \frac{\partial D}{\partial \theta_i} Z^T
\end{equation*}
for parameters of the covariance function, and $A_i = I$ for the variance parameter of a Gaussian distribution. Then, the information matrix for the covariance parameters has elements:
\begin{equation*}
    [M_{\theta}]_{i,j} = \text{tr}\left( \Sigma^{-1} A_i \Sigma^{-1} A_j \right)
\end{equation*}
and equivalently for the scale parameter such as $\sigma^2$. The partial derivatives of matrix $D$ can then readily be obtained using the auto-differentiation scheme implemented in the calculation of the elements of the matrix. 

\subsection{Small sample corrections}
The GLS standard errors with maximum likelihood estimate of the model parameters are known to exhibit a small-sample bias and underestimate the standard error when, for example, the number of higher-level groups or clusters is small. The small sample bias exists for two reasons: the standard errors fail to take into account the variability from estimating the variance parameters $\theta$ and $\phi$; and, the GLS estimator is itself a biased estimator of the variance of the fixed effect parameters. So the standard GLS standard error can underestimate the sampling error of $\beta$, resulting in overly narrow confidence intervals. There have been a range of corrections proposed in the literature, typically with regards to the degrees of freedom of reference distributions for related test statistics (e.g. \citet{Satterthwaite1946,Kenward1997}). We can also use the tools in this package to generate a bias corrected variance-covariance matrix for $\hat{\beta}$ using the approach proposed by \citet{Kenward1997}. The correction is valid for restricted maximum likelihood (REML) estimates of $\theta$ and $\phi$ as the standard maximum likelihood estimates of these parameters are biased downwards due to not accounting for the loss of degrees of freedom from estimating $\beta$ \citep{Harville1977}. At the time of writing, we have not implemented REML with the Laplace approximation methods described in the preceding section. However one can view the estimates of the variance components from the MCML algorithms as representing REML estimates, since the estimated sufficient statistics for $\theta$ do not depend on $\beta$ \citep{Laird1982}. Thus, using these estimates and to apply the small sample bias correction, we implement $M_\theta$ as:
\begin{equation*}
    [M_{\theta}]_{i,j}^{corr} = [M_{\theta}]_{i,j} - \text{tr}\left(M_\beta X^T \Sigma^{-1} A_i \Sigma^{-1} [ 2I - XM_\beta X^T\Sigma^{-1} ]A_j \Sigma^{-1}X \right)
\end{equation*}
Then, letting $K = (M_\theta^{corr})^{-1}$ and 
\begin{align*}
    P_i &= -X^T\Sigma^{-1}A_i\Sigma^{-1}X \\
    Q_{ij} &= X^T\Sigma^{-1}A_i\Sigma^{-1}A_j\Sigma^{-1}X \\
    R_{ij} &= X^T\Sigma^{-1}\frac{\partial^2 \Sigma}{\partial \theta_i \partial \theta_j} \Sigma^{-1}X
\end{align*}
the bias corrected variance covariance matrix for $\beta$ is
\begin{equation*}
    V_\beta^{corr} = M_\beta^{-1} + 2M_\beta^{-1} \left( \sum_{i=1}^r \sum_{j=1}^r K_{ij}(Q_{ij} - P_iMP_j - \frac{1}{4}R_{ij}) \right)M_\beta^{-1}
\end{equation*}
The degrees of freedom correction is also given in \citet{Kenward1997}, which is the same degrees of freedom correction originally proposed by \citet{Satterthwaite1946}. The Kenward-Roger correction, though, can \textit{over-}estimate the standard errors in very small samples in some cases. \citet{Kenward2009} proposed an improved correction for covariance functions not linear in parameters (such as the autoregressive or exponential functions). The improved correction adds an additional adjustment factor. We have implemented these corrections in the package, which can returned directly from a \code{Model} object using the \code{small_sample_correction()} member function, or they can be returned during model fitting by selecting the appropriate option for the \code{se} argument  of the \code{MCML} and \code{LA} functions. For use in other calculations one can also retrieve the first and second-order partial derivatives of $\Sigma$ with respect to the covariance function parameters using \code{partial\_sigma()}.


\subsection{Modified Box correction}
Finally, we also provide the ``modified Box correction'' for Gaussian-identity GLMMs given by \citet{Skene2010}. 

\subsection{Additional inputs and outputs}
One can specify weights for the model when generating a new \code{Model} object. For binomial models the number of trials can also be passed to the model via the \code{trials} argument, and offsets can be given in the same way via the \code{offset} argument.

The \code{MCML} and \code{LA} functions return an object of (S3) class \code{mcml} for which print and summary methods are provided. The print function returns standard regression output and basic diagnostics. We implement the conditional Akaike Information Criterion method described by \citet{Vaida2005} for GLMMs, and approximate conditional and marginal R-squared values using the method described by \citet{Nakagawa2013} for GLMMs. The object returned by \code{MCML} also contains other potentially useful output. In particular, there is often interest in the random effects themselves. As these are simulated during the MCMCML algorithm, the final set of $m$ samples are returned as a $Q \times m$ matrix facilitating further analysis.

\subsection{Prediction}
\label{sec:predict}
One can generate predictions from the model at new values of $X$, $Z$, and the variables that define the covariance function. Conditional on the estimated or simulated values of $\mathbf{u}$, the distribution of the random effects at new values is:
\begin{equation*}
    \mathbf{u}_{new} \sim N\left( D_{01}D_{00}^{-1}\mathbf{u}, D_{11}-D_{01}D_{00}^{-1}D_{01}^T \right)
\end{equation*}
where $D_{00}$ is the covariance matrix of the random effects at the observed values, $D_{11}$ is the covariance matrix at the new values, and $D_{01}$ is the covariance between the random effects at the observed and new values. Where there are multiple samples of $\mathbf{u}$, the conditional mean in averaged across them. Following a model fit, we can generate the linear predictor and mean and covariance at new locations using the member function \code{\$predict(newdata)}. This differs from the functionality of the function \code{\$fitted()}, which generates the linear predictor (with or without random effects) for the existing model at the observed data locatons, and \code{\$sim\_data()}, which simulates outcome data at the existing data values.

\section{Examples}
We provide two examples illustrating model fitting and related functionality to demonstrate how \pkg{glmmrBase} may be useful in certain statistical workflows.

\subsection{Cluster randomised trial}
\label{subsec:crct}
We consider a stepped-wedge cluster randomised trial (see \citet{Hemming2020}, for example) with six cluster sequences and seven time periods, and ten individuals per cluster period cross-sectionally sampled. All clusters start in the control state and in each period one cluster switches to the intervention state. We simulate data from a Poisson-log GLMM with cluster and cluster-period exchangeable covariance function for the example. The Poisson-log model is, for individual $i$ in cluster $j$ at time $t$:
\begin{align*}
    y_{ijt} & \sim \text{Poisson}(\lambda_{ijt}) \\
    \lambda_{ijt} &= \exp(\beta_0 d_{j} + \beta_1I(t=1) + ... + \beta_6 I(t=5) + \gamma_{jt})
\end{align*}
where $d_j$ is an indicator for treatment status, $I(.)$ is the indicator function, and $\gamma_{jt}$ is a random effect. For the data generating process we use $\beta_0 = 0.5$ and $\beta_1,...,\beta_6$ range from -1.5 to -0.3. We specify the exchangeable correlation structure $\gamma_{jt} = \gamma_{1,j} + \gamma_{2,jt}$ where $\gamma_{1,j} \sim N(0,0.3^2)$ and  $\gamma_{2,jt} \sim N(0,0.15^2)$. See \citet{Li2021} for a discussion of such models in the context of cluster randomised trials. 

We can specify the model as follows, where we have included the parameter values:
\begin{CodeChunk}
\begin{CodeInput}
data <- nelder(~ (cl(6)*t(7))>i(10))
data$int <- 0
data[data$t > data$cl,'int'] <- 1

model <- Model$new(
   formula = ~ int + factor(t)-1 + (1|gr(cl))+(1|gr(cl,t)),
   covariance = c(0.05, 0.01),
   mean = c(0.5,seq(-1.5,-0.3,by=0.2)),
   data = data,
   family = poisson())
y <- model$sim_data()
\end{CodeInput}
\end{CodeChunk}
The final line generates random simulated data.

The code to fit these models in the popular \pkg{lme4} and \pkg{glmmTMB} packages is:
\begin{CodeChunk}
\begin{CodeInput}
lme4::glmer(y ~ int + factor(t)-1+(1|cl/t), family= poisson(), data=data)

glmmTMB::glmmTMB(y ~ int + factor(t)-1+diag(1|cl/t),data=data,family=poisson())
\end{CodeInput}
\end{CodeChunk}

We fit the models in \pkg{glmmrBase} specified above using the following code for each of the fits. Note that after fitting the model, the estimated parameters are stored in the model and will be used for subsequent model fits.
\begin{CodeChunk}
\begin{CodeInput}
model$mcmc_options$samps <- 50 # set the number of samples to 50
fit3 <- model$MCML(y) 
model$mcmc_options$samps <- 500
# MCEM, L-BFGS, convergence when difference between parameter estimates < 1e-3
fit4 <- model$MCML(y,method="mcem",algo = 1,conv.criterion = 1, tol = 1e-3) 
\end{CodeInput}
\end{CodeChunk}
Note that the number of MCMC samples is relatively small for the SAEM algorithm, as samples are stored from all previous iterations. We increase the number of samples for the MCEM algorithm.

One can obtain bias-corrected standard errors for these models either by fitting the models with the \code{se="kr"} options, or retrieve the corrected standard errors after fitting using \code{model$small_sample_correction()}.

\subsection{Geospatial data analysis with approximate covariance}
As described in Section \ref{subsec:compact} and elsewhere, fitting models with large, dense covariance matrices is computationally intensive. We provide several approximations in this package to reduce fitting times. Here, we provide an example using the nearest neighbour Gaussian process, and generating predictions across an area of interest.

We generate a 600 random points in the square $[-1,1]\times[-1,1]$. We then simulate data from a Gaussian-identity model with a spatial Gaussian process with exponential covariance function:
\begin{CodeChunk}
\begin{CodeInput}
n <- 600
df <- data.frame(x = runif(n,-1,1), y = runif(n,-1,1))
sim_model <- Model$new(
  formula = ~ (1|fexp(x,y)),
  data = df,
  covariance = c(0.25,0.3),
  mean = c(0),
  family = gaussian()
)
y_sim <- sim_model$sim_data()
\end{CodeInput}
\end{CodeChunk}
We then generate a new model with nearest neighbour Gaussian process covariance, and where we name the intercept parameter \code{baseline} and include sensible starting values:
\begin{CodeChunk}
\begin{CodeInput}
analysis_model <- Model$new(
  formula = ~ (1|nngp_fexp(x,y)),
  data = df,
  family = gaussian()
)
analysis_model$set_trace(1) # for verbose output
analysis_model$mcmc_options$samps <- 50
fit_sp <- analysis_model$MCML(y_sim,algo = 2)
\end{CodeInput}
\end{CodeChunk}
We can use the model with the fitted parameters to then generate predictions across the whole area of interest, which is useful for plotting and other applications. 
\begin{CodeChunk}
\begin{CodeInput}
# a regular lattice across the area of interest
plot_data <- expand.grid(x = seq(-1,1,by=0.05), y = seq(-1,1,by=0.05)) 
# this will return a list with the linear predictor and latent effects
preds <- analysis_model$predict(newdata = plot_data) 
plot_data$value <- preds$re_parameters$vec
require(ggplot2)
ggplot(data=plot_data,aes(x=x,y=y,fill=value))+
  geom_tile()+
  scale_fill_viridis_c()
\end{CodeInput}
\end{CodeChunk}

\section{C-Optimal Experimental Designs with glmmrOptim}
The \pkg{glmmrOptim} package provides a set of algorithms and methods to solve the c-optimal experimental design problem for GLMMs. More detail on the methods in this package and a comparison of their performance can be found in \citet{Watson2022}. 

We assume there are $n$ total possible observations indexed by $i \in [1,...,n]$ whose data generating process is described by a GLMM described in Section \ref{subsec: glmm_framework}. The observations (indexes) are grouped into $J$ `units', for example clusters or individual people, which are labelled as $\mathcal{E}_j \subset [1,...,n]$. These units may contain just a single observation/index. The set of all units is the \textit{design space} $\mathcal{D}:={e_1,...,e_J}$ and a `design' is $d \subset D$. The optimal experimental design problem is then to find the design $d$ of size $J' < J$ that minimises the c-optimality function:
\begin{equation}
\label{eq:coptim}
    g(d) = c^T M_d^{-1} c
\end{equation}
where $M$ is the information matrix described in Equation \ref{eq:infomat} associated with design $d$. There is no exact solution to this problem, in part because of the complexity resulting from the non-zero correlation between observations and experimental units. 

There are several algorithms that can be used to generate approximate solutions. A full discussion of this problem in the context of GLMMs and evaluation of relevant algorithms is given in \citep{Watson2022} (see also \citet{Fedorov1972,Wynn1970,Nemhauser1978,Fisher1978}). Here, we briefly describe the relevant algorithms and their implementation in the \pkg{glmmrOptim} package.

\subsection{Existing software}
The package \pkg{glmmrOptim} provides algorithms for identifying approximate c-optimal experimental designs when the observations and experimental units may be correlated. In \proglang{R}, there are several packages that provide related functionality. The \pkg{skpr} package provides D, I, Alias, A, E, T, and G-optimal, but not c-optimal, designs for models including with correlated observations, but only where the experimental units are uncorrelated. The \proglang{R} package \pkg{AlgDesign} also provides algorithms for estimating D, A, and I-optimal experimental designs, including for models with correlated observations, but again without correlation between experimental units. \pkg{AlgDesign} implements a version of the Greedy Algorithm (described below) for D-optimality. Both packages utilise, among other approaches, the approximation described below in Section \ref{subsec:uncorexp}. Other relevant packages include \pkg{OptimalDesign}, which uses the commercial software \pkg{Gurobi}, and the packages \pkg{FrF2} and \pkg{conf.design}, which consider factorial designs specifically. Only \pkg{skpr} provides both optimal experimental design algorithms and associated methods like Monte Carlo simulation and power calculation. \pkg{glmmrBase} can provide power calculations using the functions discussed in Section \ref{subsec:power}, as well as the other suite of methods to interrogate and fit relevant models once an optimal design has been identified, which emphasizes the usefulness of the linked set of packages we present.

\subsection{Combinatorial Optimisation Algorithms}
\label{subsec:combinopt}
There are three basic combinatorial algorithms relevant to the c-optimal design problem. `Combinatorial' here means choosing $J'$ discrete elements (experimental units) from the set of $J$ to minimise Equation (\ref{eq:coptim}). 

Algorithm 3 describes the `local search algorithm'. This algorithm starts with a random design $d_0$ of size $J'$ and then makes the optimal swap between an experimental unit in the design and one not currently in the design. It proceeds until no improving swap can be made.  Algorithm 4 shows the `reverse greedy search algorithm' that instead starts from the complete design space and iteratively removes the worst experimental unit until a design of size $J'$ is achieved. A third algorithm, the `greedy search' is also included in the package, which starts from a small number of units and iteratively adds the best unit. However, this algorithm performs worse than the other two in empirical comparisons and so is not discussed further here. We offer these three algorithms in the \pkg{glmmrOptim} package. We allow the user to use any combination of these algorithms in succession. While such combinations do not necessarily have any theoretical guarantees, they may be of interest to users of the package. A more complete discussion of the methods provided by this package can be found in \citet{Watson2022}, with further examples in \citet{Watson2023}.

\begin{algorithm}
\caption{Local search algorithm}
\begin{algorithmic}
 \State Let $d_0$ be size $J'$ design 
 \State Set $\delta = 1$ and $d \leftarrow d_0$ 
 \While{$\delta > 0$}
 \ForAll{element $e_j \in d$ and $e_{j'}\in D / d$}
    Calculate $g(d / \{e_j\} \cup \{e_{j'}\})$ 
 \EndFor
 \State Set $d' \leftarrow \argmin_{j,j'} g(d / \{e_j\} \cup \{e_{j'}\})$ 
 \State $\delta = g(d') - g(d)$ 
 \If{$\delta > 0$ }
    $d \leftarrow d'$
 \EndIf
 \EndWhile
\end{algorithmic}
 \end{algorithm}

 \begin{algorithm}
\caption{Reverse greedy search algorithm}
\begin{algorithmic}
 \State Set $d \gets D$
 \State Set $k = \vert D \vert$;
 \While{$k > J'$}
 \ForAll{element $e_{j}\in d$}
    Calculate $g(d / \{e_{j}'\})$
 \EndFor
 \State Set $d \leftarrow d / \argmin_{e_j} g(d / \{e_j\})$ \;
 \State $k \leftarrow k - 1$
 \EndWhile
\end{algorithmic}
\end{algorithm}

\subsubsection{Computation}
We use several methods to improve computational time to execute these algorithms. The most computationally expensive step is the calculation of $g(d)$ after adding or swapping an experimental unit, since the calculation of the information matrix requires inversion of the covariance matrix $\Sigma$ (see Equation (\ref{eq:infomat})). However, we can avoid inverting $\Sigma$ on each step by using rank-1 up- and down-dating.

For a design $d$ with $J'$ observations with inverse covariance matrix $\Sigma^{-1}_d$ we can obtain the inverse of the covariance matrix of the design with one observation removed $d' = d / \{i\}$, $\Sigma^{-1}_{d'}$ as follows. Without loss of generality we assume that the observation to be removed is the last row/column of $\Sigma^{-1}_d$. We can write $\Sigma^{-1}_d$ as 
\begin{equation*}
    \Sigma^{-1}_d = B = \begin{pmatrix}
     C & f \\
     f^T & e \\
    \end{pmatrix}
\end{equation*}
where $C$ is the $(J'-1) \times (J'-1)$ principal submatrix of $B$, $f$ is a column vector of length $(J'-1)$ and $e$ is a scalar. Then,
\begin{equation*}
     \Sigma^{-1}_{d / \{i\}} = G =  C - ff^T/e
\end{equation*}

For a design $d$ with $J'$ observations with inverse covariance matrix $\Sigma^{-1}_d$, we aim now to obtain the inverse covariance matrix of the design $d' = d \cup \{i'\}$. Recall that $Z$ is a $n \times Q$ design effect matrix with each row corresponding to an observation. We want to generate $H^{-1} = \Sigma_{d'}^{-1}$. Note that:
\begin{equation*}
    H = \Sigma_{d'} = \begin{pmatrix}
    G^{-1} & k \\
    k^T & h \\
    \end{pmatrix}
\end{equation*}
where $k = Z_{i \in d}DZ_{i'}$ is the column vector corresponding to the elements of $\Sigma = W^{-1} + ZDZ^T$ with rows in the current design and column corresponding to $i'$, and $h$ is the scalar $W^{-1}_{i',i'} + Z_{i'}DZ_{i'}^T$. Also now define:
\begin{equation*}
    H^* = \begin{pmatrix}
    \Sigma_d & 0 \\
    0 & h 
    \end{pmatrix}
\end{equation*}
so that 
\begin{equation*}
    H^{* -1} = \begin{pmatrix}
    \Sigma^{-1}_d & 0 \\
    0 & 1/h
    \end{pmatrix}
\end{equation*}
and 
\begin{equation*}
    H^{**} = \begin{pmatrix}
    \Sigma_d & k \\
    0 & h 
    \end{pmatrix}
\end{equation*}
and $u = (k^T, 0)^T$ and $v=(0,...,0,1)^T$, both of which are length $J'$ column vectors. So we can get $H^{**}$ from $H^*$ using a rank-1 update as $H^{**} = H^* + uv^T$ and similarly $H = H^{**} + vu^T$. Using the Sherman-Morison formula:
\begin{equation*}
    H^{**-1} = H^{* -1} - \frac{H^{* -1}uv^TH^{* -1}}{1+v^TH^{* -1}u}
\end{equation*}
and 
\begin{equation*}
    H^{-1} = H^{** -1} - \frac{H^{** -1}vu^TH^{** -1}}{1+u^TH^{**-1}v}
\end{equation*}
So we have calculated the updated inverse with only matrix-vector multiplication, which has complexity $O(n^2)$ rather than the $O(n^3)$ required if we just took a new inverse of matrix $\Sigma_{d'}$.

Other steps we include to improve efficiency are to check if any experimental units are repeated, if so then we can avoid checking a swap of an experimental unit for itself. Internally, the program only stores the unique combinations of rows of $X$ and $Z$ and tracks the counts of each in or out of the design. Finally, when the function is executed (see Section \ref{subsec:implementoptim}), a check is performed to determine whether the experimental units are all uncorrelated with one another. If they are uncorrelated then we can implement an alternative approach to calculating $g(d)$, since we can write the information matrix as:
\begin{equation}
\label{eq:infomatsum}
    M_d = \sum_{j = 1}^{J'} M_{e_j} = \sum_{j=1}^{J'}X_{i \in e_j}^T\Sigma^{-1}_{i \in e_j}X_{i \in e_j}
\end{equation}
where we use $X_{i \in e_j}$ to indicate the rows of $X$ in condition $e_j$, and $\Sigma^{-1}_{i \in e_j}$ the submatrix of $\Sigma^{-1}$ in $e_j$. Thus, rather than using a rank-1 up- or down-date procedure iterated over the observations in each experimental unit, we can calculate the `marginal' information matrix associated with each experimental unit and add or subtract it from $M_d$ as required. This method is generally faster as the number of observations per experimental unit is often much smaller than $J'$.

\subsection{Approximate Unit Weights}
\label{subsec:uncorexp}
If the experimental units are uncorrelated, we can use a different method to combinatorial optimisation. We assume that all the experimental units in $D$ are unique (the software automatically removes duplicates when using this method), and we place a probability measure over them $\phi = \{(\phi_j,X_{i \in e_j},Z_{i \in e_j}): j=1,...,J, \phi_j \in [0,1]\}$ where $\sum_j \phi_j = 1$; $\phi$ is then an approximate design. The weights $\phi_j$ can be interpreted as the proportion of `effort' placed on each experimental unit. We can rewrite Equation (\ref{eq:infomatsum}) for the information matrix of this design as:
\begin{equation}
\label{eq:infomatsum2}
    M_\phi = \sum_{j=1}^{J}X_{i \in e_j}^T\Sigma^{-1}_{i \in e_j}X_{i \in e_j}\phi_j
\end{equation}
where the subscript $i\in e_j$ indicates the submatrix relating to the indexes in $e_j$. The problem is then to find the optimal design $\phi^* = \argmin_\phi c^T M_\phi c$.

\citet{Holland-Letz2011} and \citet{Sagnol2011} generalise Elfving's Theorem \citep{Elfving1952}, which provides a geometric characterisation of the c-optimal design problem in these terms, to the case of correlation within experimental units. \citet{Sagnol2011} shows that this problem can be written as a second order cone program, and as such be solved with interior point methods for conic optimisation problems. We modify this program to solve \citet{Holland-Letz2011} version of the theorem and include it in the \pkg{glmmrOptim} package, which we implement using the \pkg{CVXR} package for conic optimisation. 

\subsection{Optimal Mixed Model Weights}
The final alternative algorithm identifies the optimal mixed model weights. The best linear unbiased estimator for the linear combination $b = c^T\beta$ can be written as
\begin{align*}
    \hat{b} &= \mathbf{a}^T\mathbf{y} \\
    &= \mathbf{a'}^TL\mathbf{y} 
\end{align*}
where $\mathbf{a} = L^T\mathbf{a'}$ is a vector of weights. It can be shown, see Watson, Girling, and Hemming REF, that the experimental unit weights $\boldsymbol{\Phi}$ that minimise the bias are generated by Algorithm \ref{alg:girling}. As before these weights can then be used to generate optimal counts for each experimental unit.

\begin{algorithm}
\caption{Optimal mixed model weights for $J$ experimental units with a target total number of observations $J'$ and $\epsilon$ is the tolerance of the algorithm.}
\label{alg:girling}
\begin{algorithmic}
\Procedure{Optimal mixed model weights}{}
 \State Let $\boldsymbol{\phi} = [\phi_1,...,\phi_J]$ with $\phi_j = 1/J'$ for all $j$
 \State Set $\delta = 1$
 \While{$\delta > \epsilon$}
 \State $a \leftarrow \Sigma^{-1}X(X^T \Sigma^{-1}X)^{-1}c$
 \ForAll{$j \in \{1,...,J\}$}
    $\phi'_{j} \leftarrow \frac{\vert a_{j} \vert}{\sum_{j} \vert a_{j} \vert}$
 \EndFor
 \State $\delta \leftarrow \argmax_j \vert \phi_j - \phi'_j \vert$
 \ForAll{$j \in \{1,...,J\}$}
    $\phi_{j} \leftarrow \phi'_j$
 \EndFor
 \State $\Sigma \leftarrow W^{-1} \text{diag}(\boldsymbol{\phi}^{-1}) + ZDZ^T$ 
 \EndWhile
\EndProcedure
\end{algorithmic}
 \end{algorithm}

 \subsection{Rounding}
 The approximate weighting approaches return the set of weights $\phi_1,...,\phi_J$ corresponding to each unique experimental unit. These weights need to be converted into an exact design with integer counts of each experimental unit $n_1,...,n_J$. The problem of allocating a fixed set of $J'$ items to $J$ categories according to a set of proportions is known as the apportionment problem. There a several rounding methods used for the apportionment problem proposed by the founding fathers of the United States for determining the number of representatives from each state. \citet{PUKELSHEIM1992} show that a modified version of Adams' method is most efficient when there needs to be at least one experimental unit of each type. In other cases the methods of Hamilton, Webster, or Jefferson may be preferred. We provide the function \code{apportion()}, which generates exact designs from all these methods for a desired sample size and set of weights. The output of this function is automatically provided when the approximate weighting method is used.

\subsection{Robust c-Optimal Designs}
The preceding discussion has assumed that the correct model specification is known. However, in many scenarios there may be multiple plausible models, and a design optimal for one model or set of parameters, may perform poorly for another. Robust optimal design methods aim to produce an (approximate) optimal design over multiple candidate models. We implement a method amendable to the combinatorial algorithms described in Section \ref{subsec:combinopt} following \citet{Dette1993}. Let $\mathcal{M}$ represent a GLMM model. We assume, following \citet{Dette1993}, that the true model belongs to a class of GLMMs $\boldsymbol{\Xi} = \{\mathcal{M}_1,...,\mathcal{M}_R\}$ and we define a vector $\rho = \{\rho_1,...,\rho_R\}$, where $\rho_r \in [0,1]$ and $\sum_{r=1}^R \rho_r = 1$, which represents the prior weights or prior probabilities of each model. There are two robust c-optimality criteria we can use. The first:
\begin{equation*}
    h(d) = \sum_{r=1}^R \rho_r \log(c^T_r M^{-1}_{d,r} c_r)
\end{equation*}
was proposed by \citet{Dette1993} and \citet{Lauter1974}. The second is the weighted mean
\begin{equation*}
    h(d) = \sum_{r=1}^R \rho_r c^T_r M^{-1}_{d,r} c_r
\end{equation*}
Both criteria result in functions with the appropriate properties to ensure the local search algorithm maintains its theoretical guarantees. \citet{Dette1993} generalises Elfving's theorem to this robust criterion, however, further work is required to `doubly' generalise it to both correlated observations and robust criterion. While there may be a straightforward combination of the work of \citet{Dette1993} and that of \citet{Holland-Letz2011} and \citet{Sagnol2011}; translating this into a program for conic optimisation methods is a topic for future research.

\subsection{Implementation}
\label{subsec:implementoptim}
The \pkg{glmmrOptim} package adds an additional \code{DesignSpace} class, which references one or more \code{Model} objects. The \code{Model} objects must represent the same number of observations, with each row of $X$ and $Z$ representing an observation. The other arguments to initialise the class are (optionally) the weights on each design, units of each observation. The observations are assumed to be separate experimental units unless otherwise specified. 

The main member function of the \code{DesignSpace} class is \code{optimal}, which runs one of the  algorithms described above and returns the rows in the optimal design, or weights for each experimental unit, along with other relevant information. If the experimental units are uncorrelated with one another then the approximate weighting method is used by default; combinatorial algorithms can instead be used with the option \code{use\_combin=TRUE}. The user can run one or more of the combinatorial algorithms sequentially. The algorithms are numbered as 1 is the local search, 2 is the greedy search, and 3 is the reverse greedy search. Specifying \code{algo=1} will run the local search. Specifying, \code{algo=c(3,1)} will first run a reverse greedy search and then run a local search on the resulting design. We note that some combinations will be redundant, for example, running a greedy search after a reverse greedy search will have no effect since the resulting design will already be of size $J'$. However, some users may have interest in combining the approaches. A list (one per model in the design space) containing the vectors $c$ in Equation (\ref{eq:coptim}) must be provided. 

For certain problems, an optimal design may include all of the same value of one or more dichotomous covariates, which would result in a non-positive definite information matrix. For example, some models include adjustment for discrete time periods, but not all time periods feature in the subset of observations in the optimal design. The program checks that the information matrix is positive definite at each step, and if not, it reports which columns may be causing the failure. These columns can then be removed using the \code{rm\_cols} argument of the \code{optimal} function.

\subsection{Examples}
\begin{figure}
    \centering
    \includegraphics[width=\textwidth]{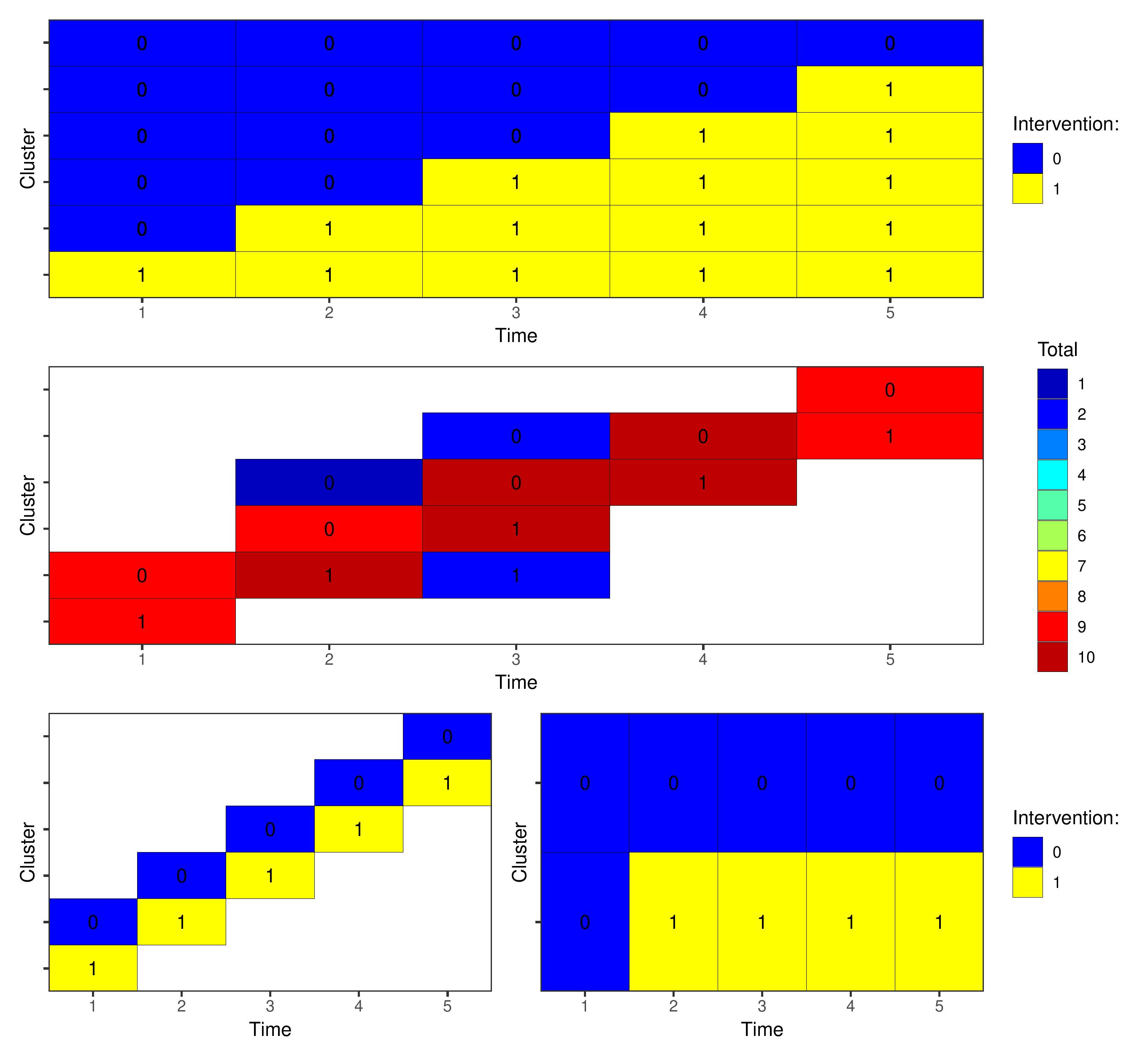}
    \caption{Top row: the cluster-trial design space. Middle: an optimal design of 100 individual observations across cluster-periods in the cluster-randomised trial design space. Bottom left: optimal design of ten complete cluster-periods comprising ten observations. Bottom right: optimal design of two cluster sequences (rows).}
    \label{fig:optim}
\end{figure}

We present a set of examples relating to identifying an optimal cluster trial design within a design space consisting of six possible sequences over five time periods shown in the top row of Figure \ref{fig:optim}. There are six clusters and five time periods, and we may observe a maximum of ten individuals per cluster-period. \citet{Girling2016} and \citet{Watson2023} consider similar design problems. 

The function that executes the algorithms in the \code{DesignSpace} class is \code{optim}. The argument \code{algo} specifies the algorithm with: 1 = local search, 2 = greedy search, 3 = reverse greedy search, and ``girling'' for the optimal mixed model weights (named after the proposer). One can also string together algorithms to run in sequence such as \code{algo = c(3,1)}.

\subsubsection{Correlated experimental units with a Single Observation}
First, we consider the case when each observation is a separate experimental unit so we may observe any number of individuals in a time period. We will identify an approximate optimal design with 100 observations with a maximum of ten observations in a single cluster-period. As the experimental units are correlated we will use the local search algorithm (option \code{algo=1}). Given that the algorithm is not guaranteed to produce the optimal design, we run it ten times and select the design with lowest value of the objective function. We will specify a Gaussian model with identity link function and exchangeable covariance function:
\begin{CodeChunk}
\begin{CodeInput}
#simulate data
df <- nelder(formula(~ (cl(6) * t(5)) > ind(10)))
df$int <- 0
df[df$t >= df$cl,'int'] <- 1
des <- Model$new(formula = ~factor(t) + int - 1+(1|gr(cl)) + (1|gr(cl,t)),
                 covariance = c(0.05,0.01),
                 mean = rep(0,6),
                 data = df,
                 family=gaussian())
ds <- DesignSpace$new(des) #create design space
opt <- ds$optimal(m=100,C=list(c(rep(0,5),1)),algo=1) #run the algorithm 
\end{CodeInput}
\end{CodeChunk}
The middle panel of Figure \ref{fig:optim} shows the approximate optimal design produced by the algorithm. Code to reproduce the plots is provided in the replication materials.

\subsubsection{Correlated experimental units with Multiple Observations}
We secondly consider the case where each cluster period is an experimental unit containing ten observations, and we aim to select a design of size ten cluster-periods.
\begin{CodeChunk}
\begin{CodeInput}
# update the experimental units
ds <- DesignSpace$new(des, experimental_condition = rep(1:30, each = 10))
opt2 <- ds$optimal(m=10,C=list(c(rep(0,5),1)),algo=1) 
\end{CodeInput}
\end{CodeChunk}
The bottom left panel of Figure \ref{fig:optim} again shows the approximate optimal design produced by the algorithm, reflecting the `staircase' design from the previous example.

\subsubsection{Uncorrelated experimental units}
Finally, we conside the case where each whole cluster represents an experimental unit and we aim to pick two of these six clusters. In this example, the experimental units are uncorrelated. By default the \code{optim} function will use the second-order cone program and return optimal weights for each experimental unit. We can force the function to instead use the local or greedy search algorithms with the option \code{force\_hill=TRUE}:
\begin{CodeChunk}
\begin{CodeInput}
# update the experimental units
ds$experimental_condition <- df$cl
opt3 <- ds$optimal(m=2,C=list(c(rep(0,5),1)),algo=1,use_combin=TRUE) 
\end{CodeInput}
\end{CodeChunk}
Figure \ref{fig:optim} again shows the approximate optimal design produced by the algorithm. We note that a design containing rows 1 and 5 achieves the same variance. We can compare these results to the approximate optimal weights:
\begin{CodeChunk}
\begin{CodeInput}
w <- ds$optimal(m=2,C=list(c(rep(0,5),1)),algo=1)
w$weights 
\end{CodeInput}
\begin{CodeOutput}
0.2419337 0.1290334 0.1290329 0.1290329 0.1290334 0.2419337
\end{CodeOutput}
\end{CodeChunk}

\section{Discussion}
In this article we describe the \pkg{glmmrBase} package for R that provides functionality for model specification, analysis, fitting, and other functions for GLMMs. We also describe the \pkg{glmmrOptim} package that extends its functionality. The number of packages in R alone that provide similar functionality attests to the growing popularity of GLMMs in statistical analyses. For example, we identified eight R packages for calculating power for cluster-randomised trials alone using similar models. Our intention with \pkg{glmmrBase} is to provide a broad set of tools and functions to support a wide range of GLMM related analyses, while maintaining a relatively simple interface through the \pkg{R6} class system. For an analysis like a power calculation, the package can provide a direct estimate, any of the intermediary steps for similar analyses, or related functionality like data simulation. The power analysis can also be applied to a design identified through the \pkg{glmmrOptim} package as classes in this package inherit the functions of the \pkg{glmmrBase} classes. This support is intended to reduce the requirement for multiple different packages for different model structures where comparisons may be required. 

The \pkg{glmmrBase} package provides MCML model fitting, along with a Laplace approximation. The package provides a wider range of covariance functions than is available in other packages in \proglang{R}, it allows for non-linear functions of parameters, and it can provide robust and bias-corrected standard errors. We plan to continue to expand this functionality and available outputs, and encourage users to suggest features. The package can therefore provide a complementary approach to GLMM model fitting alongside the existing software in \proglang{R} and other software. MCML may be especially useful when the approximations provided by packages like \pkg{lme4} or \pkg{glmmTMB} may fail to produce reliable estimates or their algorithms may not converge. For simpler models, such as those with exchangable covariance functions or large numbers of observations, methods like PQL and Laplacian approximation can produce highly reliable estimates, and in a fraction of the time that MCML can. However, in more complex cases there can be differences. More complex covariance structures are of growing interest to more accurately represent data generating processes. We have used a running example of a cluster randomised trial over multiple time periods. Early efforts at modelling used a single group-membership random effect, whereas more contemporary approaches focus on temporal decay models, like autoregressive random effects \citep{Li2021}. However, \pkg{glmmTMB} is one of very few packages, prior to \pkg{glmmrBase} to offer model fitting for this covariance structure in \proglang{R}; SAS provides similar functionality but not in conjunction with other features like non-linear fixed effect specifications.

We also discussed the \pkg{glmmrOptim} package, which provides algorithms for identifying approximate optimal experimental designs for studies with correlated observations. Again, this package is design to complement existing resources. The package was built to implement a range of relevant algorithms, and so provides the only software for c-optimal designs with correlated observations. However, for other types of optimality, including D-, A, or I-, other packages will be required, particularly \pkg{skpr}, although we are not aware of approaches that can handle these types of optimality yet for designs with correlated experimental units. One of the advantages of the integrated approach offered by our package is that one can quickly generate the design produced by the algorithm and interrogate it to calculate statistics like power, generate relevant matrices for subsequent analyses, and to simulate data.  

Planned future developments of the package include allowing for heteroskedastic models with specification of functions to describe the individual-level variance and adding more complex models such as hurdle models, mixture models, and multivariate models.


\bibliography{ref}


\end{document}